\documentclass[acmsmall, manuscript]{acmart}
\AtBeginDocument{%
  \providecommand\BibTeX{{%
    \normalfont B\kern-0.5em{\scshape i\kern-0.25em b}\kern-0.8em\TeX}}}

\setcopyright{none}
\copyrightyear{2023}
\acmYear{2023}
\usepackage{amsthm}
\usepackage[utf8]{inputenc}

\usepackage{multirow}
\usepackage{mathtools}
\usepackage{xspace}
\usepackage{hyperref}
\usepackage{dirtytalk}
\usepackage{enumitem}
\usepackage{rotating}
\usepackage{subcaption}
\usepackage{afterpage}

\usepackage{longtable}
\usepackage{array}

\newcolumntype{b}{X}
\newcolumntype{s}{>{\hsize=.5\hsize}X}
\newcolumntype{L}[1]{>{\raggedright\let\newline\\\arraybackslash\hspace{0pt}}m{#1}}
\newcolumntype{C}[1]{>{\centering\let\newline\\\arraybackslash\hspace{0pt}}m{#1}}
\newcolumntype{R}[1]{>{\raggedleft\let\newline\\\arraybackslash\hspace{0pt}}m{#1}}

\begin{document}
%%
%% The "title" command has an optional parameter,
%% allowing the author to define a "short title" to be used in page headers.
\title{Generative Models for Synthetic Urban Mobility Data: A Systematic Literature Review}

%%
%% The "author" command and its associated commands are used to define
%% the authors and their affiliations.
%% Of note is the shared affiliation of the first two authors, and the
%% "authornote" and "authornotemark" commands
%% used to denote shared contribution to the research.
\author{Alexandra Kapp}
\email{alexandra.kapp@htw-berlin.de}
\orcid{0000-0002-8348-8958}
\affiliation{
  \institution{Hochschule für Technik und Wirtschaft Berlin, University of Applied Sciences}
  \city{Berlin}
  \country{Germany}}

\author{Julia Hansmeyer}
\orcid{0000-0001-5144-4443}
\affiliation{
  \institution{Hochschule für Technik und Wirtschaft Berlin, University of Applied Sciences}  \city{Berlin}
  \country{Germany}}

\author{Helena Mihaljevi\'{c}}
\orcid{0000-0003-0782-5382}
\affiliation{
  \institution{Hochschule für Technik und Wirtschaft Berlin, University of Applied Sciences}
  \city{Berlin}
  \country{Germany}}

%%
%% By default, the full list of authors will be used in the page
%% headers. Often, this list is too long, and will overlap
%% other information printed in the page headers. This command allows
%% the author to define a more concise list
%% of authors' names for this purpose.
% \renewcommand{\shortauthors}{Trovato and Tobin, et al.}

%%
%% The abstract is a short summary of the work to be presented in the
%% article.
\begin{abstract}
Although highly valuable for a variety of applications, urban mobility data is rarely made openly available as it  contains sensitive personal information. Synthetic data aims to solve this issue by generating artificial data that resembles an original dataset in structural and statistical characteristics, but omits sensitive information. For mobility data, a large number of corresponding models have been proposed in the last decade. This systematic review provides a structured comparative overview of the current state of this heterogeneous, active field of research. A special focus is put on the applicability of the reviewed models in practice.

\end{abstract}

%%
%% The code below is generated by the tool at http://dl.acm.org/ccs.cfm.
%% Please copy and paste the code instead of the example below.
%%

%\ccsdesc[300]{General and reference~Surveys and overviews}

%%
%% Keywords. The author(s) should pick words that accurately describe
%% the work being presented. Separate the keywords with commas.

%%
%% This command processes the author and affiliation and title
%% information and builds the first part of the formatted document.
\maketitle

\section{Introduction}

Urban human mobility data is crucial for various applications, including urban planning~\cite{zhou_understanding_2018}, traffic 
management~\cite{naboulsi_large-scale_2016-1}, and smart city applications~\cite{creutzig_smart_2021}. The COVID-19 pandemic further highlighted its usefulness in pandemic analysis~\cite{aktay_google_2020, gao_mapping_2020} and simulations~\cite{pesavento_data-driven_2020}. However, there are limited openly available datasets, mainly due to privacy concerns. For instance, Culnane et al. \cite{culnane_stop_2019-1} showed recently that as few as three time-stamped locations per person were sufficient to identify most of the individuals in a dataset consisting of public transit smart card records in Melbourne provided for a hackathon. 
Another example is the publication of the New York City (NYC)  taxi dataset, in which celebrities and their travel routes could quickly be identified  \cite{trotter_public_2014}.

While aggregated data can be used for some applications, innovation is limited without access to raw data. For instance, machine learning algorithms typically require granular data. These are currently employed to develop, e.g., next-location prediction models used to provide seamless mobility offers \cite{chekol_survey_2022} or traffic mode recognition that enables more precise urban mobility analyses from Global Positioning System (GPS) data \cite{shah_processing_2017}.

So far, classical anonymization techniques for location data such as obfuscation or cloaking have not been able to sufficiently balance privacy and utility or to scale to large datasets \cite{fiore_privacy_2020}. 
The generation of synthetic data, i.e., artificial data that resembles an original dataset in structural and statistical characteristics, has the potential to overcome this issue. It is considered especially useful when data release is required 
not only for the provision of open data, but also for internal sharing, software testing, development of machine learning models, or the deployment of privacy-preserving machine learning tools \cite{jordon_synthetic_2022}. 

Synthetic data has been used in other fields, such as health and finance, for diagnostic classification~\cite{benaim_analyzing_2020, chen_synthetic_2021} and fraud detection~\cite{finance_synthetic}.
However, unlike tabular data, generating synthetic urban mobility data poses specific challenges. The sparsity and high dimensionality that result from the combination of time series with rich spatial information make it difficult to preserve complex semantic dependencies while ensuring privacy. Time-series health and finance data are also highly complex and thus cause a set of challenges \cite{finance_synthetic, health_synthetic}, though they are of different nature and thus the respective approaches cannot be transferred to mobility data.

In the past few years, a plethora of research articles addressing synthetic urban human mobility data generation has been published with currently more than 50 relevant approaches.
It is thus hardly surprising that the rapid increase in knowledge production in a multidisciplinary field at the intersection of privacy, urban mobility, and data science has led to a state of research that is difficult to survey, especially regarding the applicability of the respective methods: the success of the heterogeneous technical approaches is defined and measured differently;   
while some approaches focus on providing formal privacy guarantees, others do not include any privacy considerations; the synthesis output varies, from models directed at generating fine-granular taxi trips to those striving to produce representative motions within an entire city; some algorithms utilize solely information from the given mobility dataset, while others incorporate knowledge about human mobility behavior or additional data sources such as the road network or census data.

We systematically review research addressing the generation of synthetic mobility data to provide a structured overview and comparison of existing approaches, foster further improvement of algorithms, and evaluate the applicability of the frameworks. The latter is particularly beneficial for practitioners who strive to generate synthetic mobility data in practice. Our survey allows for an informed assessment of the usefulness of methods for specific scenarios, as synthesis methods are often constructed with certain assumptions and heavily rely on the type of dataset and (implicitly) targeted applications.

Practitioners seeking to anonymize data typically search for a synthesis method that fits their specific application scenario and dataset. The following questions, which serve as the main structure of this review, are fundamental to guide such a decision-making process: (Q1) Is the synthesis method suitable for the given data? (Q2) How does the method work? (Q3) Does the method provide a required or satisfactory  level of  (a) privacy and (b) utility for the intended use case? We thus extract relevant information on datasets the developed methods are supposed to be applied to (Q1), the underlying algorithmic approaches (Q2), privacy considerations (Q3a), and utility evaluation measures (Q3b).
In addition, we have organized the published approaches into groups based on the type of mobility they aim to address. 
For example, a 20 minutes taxi trip is considered a different type of mobility compared to the movement patterns between meaningful locations of a person over a longer period of time.
Grouping the approaches in this way allows for easier assessment of their suitability for a given application scenario. After  presenting the main aspects of each method, we compare, as far as possible, the approaches  within each of the categories.

The main contributions of this article are the following: 
(1) We systematically collect and review existing models for the generation of synthetic urban human mobility data, providing an overview of the current state of research.
(2) We group, classify and compare different approaches according to used datasets, algorithms, privacy considerations, and evaluation measures.
(3) We categorize each coded publication based on the type of mobility that it targets. We then briefly describe the reviewed models and compare the models within each group, thus disentangling this heterogeneous field in terms of practical focus.

The review shows that while a variety of different models has been proposed for the generation of synthetic mobility data, the task can still be considered far from being satisfactorily solved. A closer look reveals that the task as such is rather heterogeneous and many assumptions are made in the solution approaches that are rarely formulated explicitly. Resulting models need to be seen in specific contexts and cannot generally be applied to an arbitrary type of mobility data. We find that there is not yet enough information about the performance of different approaches in comparison to one another and that utility and privacy evaluations are generally  not yet extensive and meaningful enough to adopt models reliably in practice. Still, we find a lot of promising approaches whose potential for specific use cases needs further evaluation in real-life settings.

The remainder of this article is structured as follows: Section \ref{sec:background} introduces the basic concept of urban mobility data and synthetic data generation, summarizing existing literature reviews touching on the topic of synthetic mobility data. Section \ref{researchmethodology} presents the systematic literature review method adopted in this paper. 
In Section \ref{sec:descriptiveanalysis}, we provide a structured overview over the key aspects of synthetic data generation models. In Section \ref{sec:assessment} we describe, assess, and, if possible, compare the proposed models.
Finally, a discussion concludes the paper.

\section{Background}
\label{sec:background}

\textit{Urban human mobility} refers to the movement of people within cities and therefore focuses on a limited urban perimeter .%\cite{noauthor_urban_nodate}. 
As such, it does not include inter-urban mobility (travel from one city to another) or rural mobility, and thus usually excludes travel by trains or planes. Urban mobility research and development focus in particular on the movement and transport of people in urban spaces and look at mobility needs, services, and the urban environment from perspectives such as sustainability, accessibility, or safety. Typical data sources thus include recurrent movements of individuals that can be derived from travel card data, taxi logs, GPS-enabled tracking devices, surveys or mobile phone records. 
Accordingly, we do not subsume mobility on smaller scales such as indoor movements or pedestrian movements at crossings or subway stations under the term.

Human mobility is typically captured by trajectories. A \textit{trajectory} is commonly defined as a time-ordered sequence of spatio-temporal points, each of which consists of a timestamp or time segment and a pair of spatial coordinates in a given reference system such as latitude and longitude. However, to reduce complexity, many approaches for the generation of synthetic data omit timestamps and consider spatial points only. In practice, time-ordered spatial sequences alone can provide valuable information as well; thus, we relax the definition by referring to any dataset that contains spatial or  spatio-temporal sequences reflecting the movement of individuals in cities as an \textit{urban human mobility dataset} or, for brevity, \textit{mobility dataset}.\footnote{Note that a mobility dataset can contain additional information such as transport mode or semantic information related to the visited location.} 

Generally, \textit{staypoints} refer to  locations a person spends a significant amout of time (e.g., home, supermarket, kindergarten, work, etc.), in contrast to \textit{waypoints} that are only passed by on a route from one staypoint to another. A single taxi trip recorded by a GPS-device would thus consist of many waypoints connecting two staypoints while a mobility diary of a survey would include a sequence of staypoints. 
We refer to the different types of trajectories, like staypoint or waypoint trajectories, as \textit{trajectory semantics}.

Often the continuous geographic space of coordinates is discretized, for example by clusterings yielding a finite number of non-overlapping geographic domains, or by mapping coordinates to a pre-defined \textit{tessellation} consisting of non-overlapping polygons in the geographic space, called \textit{tiles}. Typical tessellations include uniform grids, mostly with squared or hexagonal tile shapes, or custom-defined irregular tiles like administrative boundaries or census cells.

\textit{Synthetic data} refers to artificial data generated from a real dataset using a \textit{generative model} that is fit to reproduce certain structural and statistical characteristics of the original data. 
Moreover, assuming that the models resemble certain statistical properties of the original dataset, they can be applied to generate an arbitrary amount of new trajectories. Current literature provides various approaches to developing a  generative model, from Markov models to deep learning algorithms. Our literature review will enclose all approaches in the above broad sense of a generative model that can be applied to generate a  mobility dataset (cf. \cite{jordon_synthetic_2022, 10.1145/3485125}). At the same time, we restrict to methods that create synthetic data to resemble an existing `static' dataset, in contrast to solutions for location data protection in an online location-based service application such as a restaurant recommendation smartphone app (cf. \cite{erdemir2020privacy, qiu2020mobile, wang2018artificial}). 

One of the main purposes for creating synthetic data is to preserve the privacy of individuals represented in the original data and thus enable wider access to relevant information and patterns without compromising individuals' privacy. 
It is considered one of the main privacy-enhancing approaches for the near future \cite{polonetsky_privacy_2020}. At the same time, statistical distributions of synthetic data are expected to approximate well those of the original data, thus, analyses performed on both should yield similar results. The process, called \textit{synthesis}, consists of building a model that captures structural and statistical properties such as multivariate relationships and interactions. Once a model is built, it can be used to sample or generate artificial data. The degree to which the synthetic data can accurately represent the real data is generally denoted as \textit{utility} of the respective model (for the given dataset). The synthesis is an important factor for the levels of both utility and privacy that can be achieved \cite{emam_practical_2020, noauthor_synthetic_nodate}.

Several existing surveys address research on privacy of mobility data. While some include methods for the generation of synthetic mobility data, none of them provide a comprehensive overview of the current state-of-the-art in this area which has grown remarkably in the last years.  
Errounda et al. \cite{errounda_analysis_2019} review approaches for securing differential privacy in the context of location data, with only a small portion of works dealing with the publication of private trajectory data. Fiore et al. \cite{fiore_privacy_2020} on the other hand focus on privacy-preserving  publishing of trajectories and elaborate on research about attacks  against anonymized trajectories as well as solutions proposed to protect databases from such attacks. The authors list a few models for synthetic data generation (that do not utilize machine learning). Jiang et al. \cite{jiang_location_2021} examine privacy-preserving techniques in location-based services, thus focusing on real-time data and  location-preserving privacy mechanisms based on obfuscation, cryptography and cooperation and caching. 

with synthesis methods based on generative adversarial networks (GAN), convolutional neural networks (CNN) and recurrent neural networks (RNN).  
Benarous et al. \cite{benarous_synthesis_2022-1} systematically address the trade-off between privacy and accuracy when generating synthetic data for long location sequences by evaluating basic implementations of methods based on long short-term memory networks (LSTMs), Markov chains, and variable-order Markov models. 
Shin et al. \cite{shin_user_2020}  review and summarize models for mobility data synthesis based on generative adversarial networks.

\section{Research methodology}
\label{researchmethodology}

We systematically collect and review all published research literature addressing the generation of synthetic  urban human mobility data. 
Our survey is based on the Digital Science Dimensions platform \cite{noauthor_digital_2018}, since it shows the most exhaustive journal coverage among typically utilized scholarly databases  \cite{singh_journal_2021-1,martin-martin_google_2021}. Our process steps are described below and summarized in Figure \ref{fig:reviewScheme}. The process from the search query compilation to reference checking was completed in the time frame from March to June 2022. Another search was performed in March 2023 to include literature published in the meantime.
\begin{figure}[tb]
\begin{footnotesize}
    
    \medskip
     \Description[Process of literature search]{<The process of the literature search is conduced in 5 steps: first, relevant terms for the search query have been compiled, secondly, a structured search was run with a Boolean search query. Results were restricted to only English results that have been published between 2012 and 2022, within the research fields `Information and Computing Sciences' or `Urban and Regional Planning'. Duplicates were removed from the results list. the database was accessed on 15.03.2022 and results were updated on the 20.9.2022. This resulted in 1,456 publications. As a third step, the publication titles and abstracts were screened and records without explicit reference to the generation of synthetic urban human mobility trajectory data was excluded. 94 publications remained. As a fourth step, the full text of the remaining publications were read to assess the quality and eligibility. Articles were excluded that did not propose a novel model for synthetic data generation, or that did not have urban human mobility trajectories as input data, or that did not have the format of single trajectories as output data. 29 publications remained. A forward and backward reference checking was conducted and 15 additional suitable studies that satisfied the inclusion criteria were included, resulting in a total of 44 publications.>} 
    
    \noindent\fcolorbox{black}{white}{
        \minipage[t]{\dimexpr0.78\linewidth-2\fboxsep-2\fboxrule\relax}
            \textbf{1. Search query  compilation} \newline
            Compile relevant search terms 
        \endminipage}\hfill
        \fcolorbox{white}{white}{
        \minipage[t]{\dimexpr0.2\linewidth-2\fboxsep-2\fboxrule\relax}
        \endminipage}
    \medskip
    
    \noindent\fcolorbox{black}{white}{
        \minipage[t]{\dimexpr0.78\linewidth-2\fboxsep-2\fboxrule\relax}
            \textbf{2. Structured search} \newline
            Search in title and abstract with the following query:\newline
            \textit{(mobility OR trajectory OR movement OR trip OR `sequential data') AND synth* AND generat*}
            
            Restrict to English articles, published between 2012 and 2023 within the research fields `Information and Computing Sciences' or `Urban and Regional Planning'.
            Remove duplicates from results list.            
        \endminipage}\hfill
        \fcolorbox{black}{white}{
        \minipage[t]{\dimexpr0.2\linewidth-2\fboxsep-2\fboxrule\relax}
             1,775 publications
        \endminipage}
    
    \medskip
    
    \noindent\fcolorbox{black}{white}{
        \minipage[t]{\dimexpr0.78\linewidth-2\fboxsep-2\fboxrule\relax}
            \textbf{3. Screening of titles and abstracts} \newline 
            exclude records without explicit reference to generation of synthetic urban human mobility trajectory data
        \endminipage}\hfill
        \fcolorbox{black}{white}{
        \minipage[t]{\dimexpr0.2\linewidth-2\fboxsep-2\fboxrule\relax}
            115 publications 
        \endminipage}
    
    \medskip
    
    \noindent\fcolorbox{black}{white}{
        \minipage[t]{\dimexpr0.78\linewidth-2\fboxsep-2\fboxrule\relax}
            \textbf{4. Quality and eligibility assessment} \newline
            Exclude articles that do not propose a novel model for synthetic data generation; or do not have urban human mobility trajectories as input data; or do not have the format of single trajectories as output data
        \endminipage}\hfill
        \fcolorbox{black}{white}{
        \minipage[t]{\dimexpr0.2\linewidth-2\fboxsep-2\fboxrule\relax}
            34 publications
        \endminipage}
        
    \medskip
    
    \noindent\fcolorbox{black}{white}{
        \minipage[t]{\dimexpr0.78\linewidth-2\fboxsep-2\fboxrule\relax}
            \textbf{5. Reference checking} \newline
            Include relevant studies from forward and backward references.
            Forward checking: examine  publications that have cited the studies (Google Scholar).
            Backward checking: examine references of the studies
        \endminipage}\hfill
        \fcolorbox{black}{white}{
        \minipage[t]{\dimexpr0.2\linewidth-2\fboxsep-2\fboxrule\relax}
            (+17 publications) \newline
            51 publications
        \endminipage}
    \caption{Overview of all steps of the literature search and resulting number of included publications.}
    \label{fig:reviewScheme}
\end{footnotesize}
\end{figure}

We initially performed a search with the rather broad OR-query \textit{synthetic trajectory generation} to determine a set of relevant search terms and their synonyms. From this, we developed the following Boolean search query: \textit{(mobility OR trajectory OR movement OR trip OR `sequential data') AND synth* AND generat*}\footnote{For details on search query syntax in the Dimensions platform see \url{https://dimensions.freshdesk.com/support/solutions/articles/23000018802-how-to-search-in-dimensions}}. This query was applied to the title and abstract fields and we limited the time range to the years 2012 to 2023.
In addition, we limited the search to articles in English language, published in the research fields  \textit{Information and Computing Sciences} or \textit{Urban and Regional Planning}.
The respective structured search resulted in 2,035 articles, with 1,775 remaining after deduplication. 
The screening of titles and abstracts of these articles was performed by the first two authors. Every record without an explicit reference to the generation of synthetic urban human mobility data was excluded. 
A parallel independent assessment of the screening procedure with a random test set comprising 140 articles ($\sim$10\% of all articles) was performed by both researchers, yielding a Cohen's Kappa value of 0.925 and thus a very high inter-rater reliability. Any remaining discrepancies on the exclusion of papers were discussed and resolved. 
This step resulted in 115 articles included for an evaluation of their full texts.

The full texts of the remaining articles were skimmed through to further asses the quality and eligibility of the presented studies. The task was split between all three authors and performed independently. 
A total of 34 articles remained based on the following inclusion criteria:
(1) The publication must propose a  \textbf{generative synthesis model} as defined in Section \ref{sec:background}. 
(2) The model must use and produce \textbf{urban human mobility data} (see Section \ref{sec:background}). This excludes papers that solely or primarily address other types of movements such as robotic movements, indoor mobility, or animal traces.
(3) Input and output data need to have the \textbf{format of single trajectories}, i.e., consist of sequences of spatial or spatio-temporal points (see Section \ref{sec:background}). (In particular, we excluded papers that generate synthetic mobility networks, e.g., \cite{mauro_generating_2022-1}.)
(4) The publication needs to \textbf{propose a novel model}. We excluded papers that solely apply or compare existing models, or vision papers that have not explicitly formulated their method (e.g., \cite{liu2018trajgans}).

To achieve a complete account of the research field, forward and backward references were examined for additional relevant literature. Google Scholar\footnote{\url{www.scholar.google.com}} was utilized for forward checking of references. The reference list of each paper was consulted for backward checking. Overall, 17 additional studies were identified and added to our literature collection, 
yielding a total of 51 coded publications  \cite{10.1007/978-3-031-13448-7_7, anda_synthesising_2021, badu-marf_composite_2020, berke_generating_2022, bindschaedler_synthesizing_2016-1, blanco-justicia_generation_2022, bwambale_getting_2021, cao2021generating, chen_differentially_2012-3, chen2012differentially, chen2020rnn, choi_trajgail_2021, dandekar_trajectory_2016, deldar_enhancing_2020, feng_learning_2020, ghane_tgm_2020, gursoy2020utility, gursoy_differentially_2019-1, gursoy_utility-aware_2018-1, he_dpt_2015-1, inproceedings, kang_trag_2021,  kulkarni_generating_2017-1, kulkarni_generative_2018-1,lestyan_search_2022, li_fts_2016, li2016differentially, liu_adaptive_2019, mir2013dp, ouyang_non-parametric_2018-1, pang_development_2020, pappalardo_data-driven_2018-1,  rao_lstm-trajgan_2020-1, roy_practical_2016, sakuma2021trajectory, smolak_population_2020,tamura_synthetic_2022, wang_large_2021, wei2020we, xusimulating, yu_reconciling_2019, zhan2022privacy, zhao_synthesizing_2019, zhou_toward_2021-1, wang_privtrace_2022, du_ldptrace_2023, sun_synthesizing_2023, jiang_continuous_2023, xiong_trajsgan_2023, yang_2022, chiesa_2022}.

\section{Structured overview}
\label{sec:descriptiveanalysis}

As outlined in the introduction, one of our main objectives is to facilitate the decision-making process for practitioners when selecting synthetic data models. This requires addressing inquiries regarding the suitability of one's data, understanding how  a method works, and determining whether the method ensures a satisfactory level of both (a) privacy and (b) utility.  

From each coded publication, we thus extracted information on the following topics: (1) utilized datasets; 
(2) algorithmic approach of the generative model; (3) privacy guarantees and evaluations; and (4) measures used to evaluate the utility.

\subsection{Datasets}
\label{sec:datasets}

The datasets used in the reviewed literature constitute `urban mobility data’ as they reveal urban movements of humans (cf. Section \ref{sec:background}). However, they differ substantially, with inherently different characteristics regarding, e.g., the typical level of (geo-temporal) granularity, additional attributes, such as demographics or traffic mode, ways they are recorded or the typically associated use cases.
Nevertheless, rather few papers specify formal requirements for the input data or explicitly state for what kind of data their method is suitable. This information is often only to be implied from the datasets chosen for evaluation or from the method description. At the same time, the properties of the datasets are paramount from an application perspective, because, for most practitioners in search of a suitable method, the dataset to be synthesized is already fixed. 

Table \ref{table:datasets} provides an overview of all mobility datasets used for evaluation in the reviewed literature including information on the dataset size, time range, and information on its availability. The indication of the dataset size is either given as the number of users, number of trajectories, or number of records; some provide a combination, some none of these. We categorize the datasets into \textit{GPS traces}, \textit{social media data}, \textit{mobile phone data}, \textit{smart card data}, \textit{surveys} and \textit{simulation datasets}. 
For detailed information on the general peculiarities of different data sources, we refer to Luca et al. \cite{10.1145/3485125}.

The heterogeneity of the size and range of datasets is striking: 
The user size ranges from 100 to almost 7 million and the time range goes from one day to 4.5 years. 
Even though the amount of data is a relevant factor for the performance of a model, almost none of the publications indicates how the dataset size relates to the utility of their model. Overall, there is little discussion of the types of datasets and properties such as the number of trips per user or temporal resolution for which the proposed procedure is (not) well suited.

 \renewcommand*{\arraystretch}{1}
  \afterpage{
 \begin{footnotesize}
\begin{longtable}{p{1mm}|p{1mm}L{32mm}R{26mm}R{14mm}R{12mm}R{16mm}R{8mm}}
	 	\caption{Datasets used for model evaluations, including size and time range. The dataset size is either provided as the number of users, number of trajectories, or number of records, i.e., single spatial points. 
		\newline * Dataset is publically available; (*) dataset is available on request. Links are provided in the supplemental materials.	\label{table:datasets}} \\
		 \hline
	\centering
 & & \textbf{Dataset name}  & \textbf{Used by} & \textbf{Users} & \textbf{Traj.} & \textbf{Records} & \textbf{Time range}  \\
\hline

\multirow{30}{=}{\rotatebox[origin=c]{90}{\textbf{GPS Traces}}} &

\multirow{3}{=}{\rotatebox[origin=c]{90}{\textbf{Bike}}}
& Bike NYC* & \cite{kang_trag_2021} & - & - &	mthly update & - \\

&  & Bike Washington*  & \cite{kang_trag_2021} &-&-& daily update &-\\  

&  & Boston Hubway Bike*  & \cite{roy_practical_2016} & -& 1 M	& - & 1 mth \\

\cmidrule{2-8}

& \multirow{3}{=}{\rotatebox[origin=c]{90}{\textbf{Car}}}
& Private Vehicle (Tuscany)&	\cite{pappalardo_data-driven_2018-1} & 159,000 & 9.8 M & - & 1 mth \\

&  & Route City (Hanghzhou)   & \cite{wei2020we} & - & - & - & - \\

& & Navigation Japan & \cite{inproceedings} & - & - & - & 1 mth \\

& & Ucar (Beijing) & \cite{sun_synthesizing_2023} & - & - & - & - \\

& & OCTO Telematics (Rome) & \cite{chiesa_2022} & 150,000 cars & - & - & 1 mth \\

\cmidrule{2-8}

& \multirow{9}{=}{\rotatebox[origin=c]{90}{\textbf{Smartphone App}}}
&  	Dino Fun World   & \cite{wei2020we} &-	&-&-&-\\

& & PRIVA' MOV (Lyon) \cite{mokhtar_privamov_2017}  & \cite{zhan2022privacy} & 100 &	80,000 & -& 1.5 yr \\

&  & Yahoo Japan disaster alert app  & \cite{pang_development_2020} & 105,200 & - & - & - \\

& & Agoop Saitama  & \cite{sakuma2021trajectory} & 5,000 & - & 3 M & 8 mth \\

&  & Blogwatcher (Nisshin City)   & \cite{tamura_synthetic_2022} & 2,155 & - & - & 1 mth \\

& & Boston LBS   & \cite{berke_generating_2022} & 22,707 & - & - & 5 d \\

& & GeoLife (Beijing)* \cite{zheng2010geolife}& \cite{gursoy2020utility, gursoy_utility-aware_2018-1, gursoy_differentially_2019-1, zhan2022privacy, feng_learning_2020, xusimulating, lestyan_search_2022, deldar_enhancing_2020, qiu2020mobile, yu_reconciling_2019, blanco-justicia_generation_2022, 10.1007/978-3-031-13448-7_7, wang_privtrace_2022} & 182 & 17,621 & - & 5 yr \\

& & MTL Trajet Study* & \cite{xiong_trajsgan_2023} & & 34,905& & 1 mth \\

\cmidrule{2-8}

& \multirow{10}{=}{\rotatebox[origin=c]{90}{\textbf{Taxi}}}
&  Beijing Taxi (2009)  \cite{lian2018one}  & \cite{deldar_enhancing_2020, he_dpt_2015-1}&	- & 1.7/4.3 M & 129 M & 	9 d \\

&  & Beijing Taxi (2013)   & \cite{li_fts_2016} & 1,725 taxis & 13,245 & 1.4 M & 1 wk \\

&  & Beijing Taxi (2015) & \cite{jiang_continuous_2023} & 15,642 & 956,070 & - & 1 wk \\

&  & Porto Taxi* \cite{moreira-matias_predicting_2013}  & \cite{gursoy_utility-aware_2018-1, gursoy_differentially_2019-1, gursoy2020utility, ghane_tgm_2020, lestyan_search_2022, wang_large_2021, deldar_enhancing_2020, du_ldptrace_2023, wang_privtrace_2022, jiang_continuous_2023, yang_2022} & 442 taxis & 1.7 M & -	 & 1 yr \\

&  & San Francisco Taxi * \cite{epfl-mobility-20090224} & \cite{kang_trag_2021, lestyan_search_2022, blanco-justicia_generation_2022} & 536 taxis & 276,744 & - & 1 mth \\

&  & Seoul Taxi  & \cite{choi_trajgail_2021} & - & 59,553 & - & 1 d \\

& & Shenzhen Taxi   & \cite{kang_trag_2021}& - & - & - & 4 d \\

& & Singapore Taxi   & \cite{cao2021generating} & 17,610 & - & ``mils. of recs.'' & 4 mth \\

& & Microsoft T-Drive (Beijing 2011)* \cite{yuan_driving_2011, yuan_t-drive_2010}  & \cite{chen2020rnn, li2016differentially} & 33,000 & 4.96 M & 790 M & 3 mth \\

& & Hangzhou Taxi  & \cite{du_ldptrace_2023} & - & 348,144 & - & - \\
& & Didi Chuxing Taxi & \cite{sun_synthesizing_2023} & - & 50,000 & - & -\\ 

\cmidrule{2-8}
\scriptsize
& \multirow{2}{=}{\rotatebox[origin=c]{90}{\textbf{Other}}}
& Kingdom of Fife  & \cite{smolak_population_2020} & 173 & - & 3,867,918 & 1 wk \\

& & LDCC \cite{kiukkonen2010towards}  (*) & \cite{kulkarni_generating_2017-1, kulkarni_generative_2018-1, bindschaedler_synthesizing_2016-1, zhan2022privacy, ouyang_non-parametric_2018-1} & 168 & - & 70 M & 1 yr \\

\hline

\multirow{7}{=}{\rotatebox[origin=c]{90}{\textbf{Social Media}}}
& & Brightkite* \cite{cho2011friendship} & 	\cite{zhou_toward_2021-1} & 58,000 & - & 4.5 M & 2.5 yr \\

& & Gowalla* \cite{cho2011friendship}  &\cite{zhou_toward_2021-1, yang_2022} & 196,591 & - & 6 M & 1.5 yr \\ 

& & NYC Restaurant Rich Dataset* \cite{yang_fine-grained_2013, yang_sentiment-enhanced_2013, yang_sesame_2014} & \cite{zhou_toward_2021-1} & 3,112 & - & 27,149 & 4 mth \\

& & NYC\&Tokyo Dataset* \cite{yang_modeling_2015}& \cite{zhou_toward_2021-1, zhan2022privacy} & - & - & 800,000 & 10 mth \\

& & Foursquare Weekly* \cite{may_petry_marc_2020} & \cite{rao_lstm-trajgan_2020-1} & 193 & 3,079 & 66,962 &-\\

\hline
\multirow{6}{=}{\rotatebox[origin=c]{90}{\textbf{Mobile Phone}}}
& & CDR (Dhaka)   & \cite{bwambale_getting_2021} & 6.9 M & - & 600 M & 2 wk \\

& & CDR (Europe)   & \cite{pappalardo_data-driven_2018-1} & 1 M & -& -& 3 mth \\

& & NYC CDR 2011  & \cite{mir2013dp} & 250,000 & - & $>$1 B & 3 mth \\

& & Mobile Operator China (2016)   & \cite{feng_learning_2020} & 100,000 & - & - & 1 wk \\

& & Mobile Phone (Singapore)  & \cite{anda_synthesising_2021} & 2.8 M & - & - & 1 day \\

\hline

\multirow{2}{=}{\rotatebox[origin=c]{90}{\textbf{S.C.}}}
& & STM Montreal  & \cite{chen_differentially_2012-3, chen2012differentially} & - & - & 1,210,096 & - \\

& & Smart Card Dataset   & \cite{dandekar_trajectory_2016} & 3.5 M & - & - & 1 mth \\

\hline

\multirow{2}{=}{\rotatebox[origin=c]{90}{\textbf{Sim.}}}
& & Brinkhoff (Oldenburg) (*)	\cite{brinkhoff2002framework}	 & \cite{gursoy2020utility, gursoy_utility-aware_2018-1, gursoy_differentially_2019-1, he_dpt_2015-1, liu_adaptive_2019, yu_reconciling_2019, du_ldptrace_2023, wang_privtrace_2022} & - & 50,000 & - & - \\

& & Melbourne Car \cite{mokbel2013mntg} & \cite{ghane_tgm_2020} & -& - & 10 M & - \\

& & Campus  & \cite{du_ldptrace_2023} &  & 1,000,000 &  &  \\

\hline

\multirow{2}{=}{\rotatebox[origin=c]{90}{\textbf{Surv.}}}
& & Survey Montreal* &	\cite{badu-marf_composite_2020} & 188,700 & 410,800& - & 3 mth \\
%& &  &  & & & &  & & &  &  \\
\end{longtable}
\end{footnotesize}

}

\textbf{GPS traces.}
The GPS \cite{united_states_departement_of_defense_global_2008} makes use of satellites to determine the geo-spatial positioning of electronic receivers, nowadays built into a wide range of devices such as smartphones, cars, or shared mobility vehicles.
The majority of datasets used in the reviewed publications are GPS related, originating from trips of shared bicycles, taxis, private vehicles, or smartphone apps.
A wide range of taxi datasets was used, though it should be noted that they may vary greatly, as their size ranges from 60,000 \cite{choi_trajgail_2021} to 5 million trajectories \cite{yuan_driving_2011, yuan_t-drive_2010}.
The most commonly used datasets collected with mobile phones are \textit{GeoLife (Beijing)} \cite{zheng2010geolife} and the \textit{Lausanne Data Collection Campaign (LDCC)} \cite{kiukkonen2010towards}.
GeoLife can be freely downloaded, thus its high accessibility explains its popularity.
It only includes 182 users who were  tracked over more than five years, but contributed trajectories rather sparsely,  thus yielding a total of `only' 17,621 trajectories.
LDCC, on the other hand, is based on the continuous tracking of 168 individuals over a year using mobile phones with embedded sensors (including GPS, cellular network information, and WiFi access points) to collect large quantities of continuous data pertaining to the behavior of individuals and their social networks. 
\textit{MTL Trajet Study} is another open dataset, which was collected with a smartphone app in the context of a study in the city of Montréal. It continuous tracking over the course of a month with additional information on reason for each trip and the mode of transportation provided by the participants.

\textbf{Mobile phone data.}
Telecommunication companies collect Call Detail Records (CDR) mostly for billing purposes. These records contain information on time and location (the position of the cell tower the user was connected to) during a phone call or any other billable telecommunication transaction such as sending or receiving text messages. 
None of the CDR datasets used in the reviewed articles is publicly available; instead, they have only been shared with the respective researchers directly by the providers. The lack of data management plans and information on the data source inhibits the reproducibility of the respective studies \cite{10.1145/3485125}.

\textbf{Social media data.}
Location-based social networks (LBSN) reflect real-life social networks through websites or mobile phone applications where users can share ideas, photos, activities, events, and interests. 
Such LBSN provide either a precise geo-tag or the position of a predefined location, like a city, an area, or a restaurant. Based on a sequence of published posts by a user, the spatial and temporal information can be used to construct a user's trajectory \cite{10.1145/3485125, zheng_location-based_2011}.
There are three LBSN datasets from the platforms Foursquare, Gowalla, and Brightkite, the latter two being, meanwhile closed, location-based friendship networks where users shared their location by checking in. The corresponding datasets 
\textit{Gowalla} \cite{cho2011friendship} and \textit{Brightkite} \cite{cho2011friendship}consisting of  $\sim\!6$ million and $\sim\!4$ million user check-ins, respectively, were made publicly available through third parties who collected the data via publicly available application programming interfaces (API). 
Foursquare is a recommendation app for restaurants, stores, and other points of interest, based on users' current location and historic preferences.
Each check-in is associated with a timestamp, GPS coordinates, and a semantic meaning which is represented by fine-grained venue categories.
As Foursquare did not provide a public API, check-ins that were also shared on Twitter were retrieved, yielding three Foursquare datasets: 
(1) the \textit{NYC Restaurant Rich dataset} \cite{yang_sentiment-enhanced_2013, yang_fine-grained_2013, yang_sesame_2014}, 
(2) the \textit{NYC and Tokyo check-in dataset} \cite{yang_modeling_2015} which includes check-ins to various types of locations in NYC and Tokyo, and (3) the \textit{Foursquare Weekly} dataset \cite{may_petry_marc_2020} which is a subset of the latter and comprises only 193 users and $\sim\! 3,000$ trajectories in NYC.

\textbf{Smart card (SC) data.}
Datasets on public transportation are typically collected by operators and local government bodies for a range of commercial and planning purposes. The locations are usually limited to those of bus and metro stops and the records commonly consist of check-ins and -outs using smart cards at the first and last stop of a public transport trip \cite{pelletier_smart_2011}.
The \textit{STM Montreal} dataset \cite{noauthor_societe_nodate} consists of 1.2 million records in the Montreal transportation system. The \textit{Smart Card dataset} comprises tappings in a non-specified city of around 3.5 million commuters over a period of one month.

\textbf{Simulation (Sim.) datasets.}
To analyze the traffic of urban mobility, traffic simulation software has been widely used in research.
Even though simulated datasets likely do not come with privacy concerns, they can be useful for the evaluation of synthetic data generation models.
The \textit{Brinkhoff} dataset \cite{brinkhoff2002framework} (developed by Brinkhoff) is a simulated dataset of 50,000 trajectories in the city of Oldenburg, Germany based on a network-based generator of moving objects. It is available on request.
\textit{Melbourne car} is a simulated dataset generated by the Minnesota Web-based Traffic Generator \cite{mokbel2013mntg}, containing 10 million records within the city of Melbourne, Australia. \textit{Campus} is a synthetically generated dataeset based on buildings of British Columbia campus.

\textbf{Surveys (Surv.).}
Household surveys collect mobility data by surveying a representative group of individuals about conducted trips.
The \textit{Survey Montréal} from 2013 contains travel diaries of  $\sim\! 140,000$ individuals, including socio-economic variables such as age, employment status, gender, etc., and other trip-related variables such as origin and sequence of trip destinations.

\subsection{Algorithms}
\label{sec:algorithms}
The `traditional' approaches to generating synthetic mobility data are based on extracting and combining  particular mobility aspects from data to generate time- and location-based components of trajectories. This is often achieved by approximating distributions of trip lengths, speed, start locations, start times, or start-end-locations \cite{gursoy_utility-aware_2018-1, li_fts_2016,liu_adaptive_2019, roy_practical_2016, sun_synthesizing_2023, mir2013dp} and  drawing samples from those. For instance, the beginning element of a trajectory can be generated by drawing independent random samples from the distribution of start locations and start times, respectively. In order to preserve a certain level of mobility behavior in the synthetic data, transition probabilities (and other information) are often estimated with, e.g., Markov chains, higher-order Markov models (computing $n$-grams and prefix trees) or hidden Markov models \cite{deldar_enhancing_2020,gursoy_differentially_2019-1,chen_differentially_2012-3,he_dpt_2015-1,ghane_tgm_2020, du_ldptrace_2023, wang_privtrace_2022, yang_2022}. Typically, all such `model-based'  approaches are rather complex, making assumptions on types and semantics of human mobility \cite{bindschaedler_synthesizing_2016-1,kang_trag_2021} and extracting and combining different kinds of information from the data, typically utilizing methods such as sampling, clustering, and addition of noise in order to achieve a certain level of privacy \cite{bindschaedler_synthesizing_2016-1,smolak_population_2020,li2016differentially,dandekar_trajectory_2016,he_dpt_2015-1, sun_synthesizing_2023, wang_privtrace_2022}.

Viewed over time, the first deep learning architectures are appearing as modeling approaches from 2017 onwards, increasingly replacing the hitherto predominant statistical and probability-theoretical models. The utilized deep learning architectures are mainly inspired by existing approaches to generate sequential data from human language such as sentences of documents. RNN architectures like LSTMs \cite{hochreiter1997long}, well known for training language models and thus able to predict probable words or characters given a sub-sequence,
have been adopted by considering locations analogous to words and, correspondingly, trajectories to sentences
\cite{berke_generating_2022,kulkarni_generating_2017-1,chen2020rnn,yu_reconciling_2019, blanco-justicia_generation_2022}.
The generation of new sequences then starts by providing context consisting of one or more locations to predict the next one in the sequence, and then iteratively expanding the sequence by adding newly generated synthetic locations to the context for the next location generation. The model generates a sequence until, for example, an `end of trajectory' character has been produced or a pre-defined length has been achieved.

GANs are probably an even more obvious architecture for synthesizing trajectories. A GAN consists of two networks, a generator $G$ and a discriminator $D$. The generator attempts to learn the distribution of the data, while the discriminator learns to decide whether a sample record comes from the training set or was produced by $G$. During the training, $G$ should thus learn to maximize the probability that $D$ makes a mistake. Synthetic data is then produced by the trained generator that has learned an approximation of the trajectory distribution \cite{goodfellow2014generative}.

Other generative approaches such as (variational) autoencoders are also found in the coded literature \cite{sakuma2021trajectory, lestyan_search_2022, inproceedings, zhou_toward_2021-1,zhan2022privacy, chiesa_2022}. These types of neural networks learn to encode a set of data using a compressed representation (as latent vectors or parameters of a pre-specified distribution in the latent space) from which the original data can be reconstructed without losing too much important information (and thus reducing noise when learning the lower-dimensional representation). 

Various combinations of architectures can be found in attempts to synthesize movement data. TrajGen, for instance, combines a GAN and a Seq2Seq model \cite{cao2021generating}; GANs and autoencoders are built using recurrent  \cite{rao_lstm-trajgan_2020-1,badu-marf_composite_2020,zhan2022privacy,xusimulating,wang_large_2021,jiang_continuous_2023}, convolutional \cite{cao2021generating,feng_learning_2020,ouyang_non-parametric_2018-1, xiong_trajsgan_2023} layers. 
Other approaches utilize ideas from reinforcement learning \cite{pang_development_2020,erdemir2020privacy} and combine them with GANs \cite{wei2020we,choi_trajgail_2021}.

Since most approaches are inspired by existing architectures developed for problems on text or image data, they require certain preprocessing steps to match data format requirements. The main differences occur in (1) the handling of spatial points and (2) the treatment of time information.

\textbf{Spatial information.}
Some approaches use coordinates (latitude and longitude) directly as numerical input variables \cite{sakuma2021trajectory, rao_lstm-trajgan_2020-1}, though most apply a discretization technique. 
The latter is used to reduce dimensionality and thus complexity of the geo-space or to achieve a setting analogous to  language models, where words are discrete objects from an existing vocabulary. 
Commonly, coordinates are mapped to tiles of a given tessellation \cite{berke_generating_2022, kulkarni_generative_2018-1, gursoy_differentially_2019-1, he_dpt_2015-1, ghane_tgm_2020, badu-marf_composite_2020, mir2013dp, smolak_population_2020, feng_learning_2020, xu_trajectory_2017, ouyang_non-parametric_2018-1, lestyan_search_2022, tamura_synthetic_2022, kang_trag_2021, bwambale_getting_2021, anda_synthesising_2021, pappalardo_data-driven_2018-1, deldar_enhancing_2020, pang_development_2020, yu_reconciling_2019, li2016differentially, blanco-justicia_generation_2022, du_ldptrace_2023, sun_synthesizing_2023,wang_privtrace_2022}. 
Alternatively, coordinates are discretized by using a fixed number of pre-defined locations (e.g., subway stations), \cite{chen_differentially_2012-3, chen2012differentially, dandekar_trajectory_2016, roy_practical_2016, zhou_toward_2021-1}, graph representations \cite{choi_trajgail_2021, liu_adaptive_2019, jiang_continuous_2023}, or clustering \cite{li2016differentially, chen2020rnn, bindschaedler_synthesizing_2016-1}. 
While most discretization approaches omit the spatial relation between locations, \cite{wei2020we, choi_trajgail_2021} require a spatial format that can be linked to an action (e.g., turn right or left) and \cite{ghane_tgm_2020, wei2020we} make use of the information of neighboring tiles.
To maintain the spatial relation, trajectories are also modeled as images \cite{ouyang_non-parametric_2018-1, cao2021generating, xiong_trajsgan_2023}.

\textbf{Temporal information.}
Many approaches reduce their complexity by omitting the time aspect and considering mobility data solely as a sequence of geo-locations  \cite{gursoy_utility-aware_2018-1, gursoy_differentially_2019-1, kulkarni_generating_2017-1, kulkarni_generative_2018-1, chen2012differentially, chen_differentially_2012-3, he_dpt_2015-1, ghane_tgm_2020, badu-marf_composite_2020, wang_large_2021, kang_trag_2021, choi_trajgail_2021, liu_adaptive_2019, yu_reconciling_2019, dandekar_trajectory_2016, blanco-justicia_generation_2022, du_ldptrace_2023, wang_privtrace_2022, sun_synthesizing_2023}. This removes an important aspect of mobility data as temporal patterns can be highly relevant, for example, the information at what time a location is visited most or if street segments are jammed at certain times. Approaches that include a notion of time usually use one of three options: (1) For each user a location is provided for each fixed time interval, like every hour \cite{bindschaedler_synthesizing_2016-1, berke_generating_2022, feng_learning_2020, tamura_synthetic_2022, pappalardo_data-driven_2018-1, pang_development_2020, zhao_synthesizing_2019} (this is only possible if there is continuous data of each user that allows such an assignment); (2) a start time is assigned to each trajectory and, if timestamps are created for consecutive points, a constant sampling rate is used \cite{cao2021generating, deldar_enhancing_2020, lestyan_search_2022}; (3) a timestamp is modeled for each point within a trajectory, mostly reduced to the hour of day \cite{10.1007/978-3-031-13448-7_7, mir2013dp, ouyang_non-parametric_2018-1, rao_lstm-trajgan_2020-1, xusimulating, anda_synthesising_2021, deldar_enhancing_2020, roy_practical_2016, li2016differentially, sakuma2021trajectory, smolak_population_2020}. Some models also consider the stay duration at a location as temporal information \cite{ghane_tgm_2020, wei2020we,anda_synthesising_2021}.

\subsection{Privacy}
\label{sec:privacy}

The release of fine-grained mobility data presents significant privacy concerns, as evidenced by numerous scientific and journalistic investigations (cf., e.g., \cite{culnane_stop_2019-1, trotter_public_2014, de_montjoye_unique_2013}).
It thus appears not too surprising that, of the 51 encoded publications, 43 explicitly mention privacy as a motivation for synthesizing mobility data. In fact, in the majority of the coded literature, privacy serves as the primary or sole motivation. Additionally, it is argued that synthetic data can be used to augment small existing real datasets, e.g., to improve training and evaluation of machine learning models \cite{choi_trajgail_2021, inproceedings}.
Those not mentioning privacy, stem mainly from the field of urban planning, which initially employed mobility models based on predefined rules rather than real-world observations to simulate human movement \cite{10.1145/2840722}. Over time, these models have become more sophisticated and are now incorporating real-life data to generate more realistic mobility patterns that can be used for simulations and what-if analyses \cite{kopp_evaluation_2014}. For these models \cite{pang_development_2020, dandekar_trajectory_2016, wei2020we, pappalardo_data-driven_2018-1, bwambale_getting_2021, kang_trag_2021, badu-marf_composite_2020, ouyang_non-parametric_2018-1}, the main concern is how well the generated data represents a greater population.
However, overall, privacy is still the main motivation for the generation of synthetic mobility data and around half of the encoded papers include a formal privacy guarantee or a privacy evaluation, which we will summarize within this chapter.

A formal privacy guarantee provides some level of certainty that the synthetic dataset differs from the original data to a given degree and does not reproduce the original data too precisely. 
Privacy evaluations can be used alternatively or in addition to privacy guarantees to examine to what extent models preserve privacy under certain types of attacks.

\textbf{Privacy guarantees.} \textit{Differential privacy (DP)} \cite{dwork_algorithmic_2013} is almost exclusively used among the coded publications if 
a guarantee is provided. 
Broadly speaking, differential privacy guarantees that the output of an
algorithm remains nearly unchanged if the data of one individual is removed or added to the dataset.
In this way, differential privacy limits the impact of a single individual on
the algorithm's outcome, preventing the reconstruction of an individual's data.
Differential privacy comes with a parameter, usually denoted by $\varepsilon$, which defines the `privacy budget' and captures the extent of potential privacy leakage. A typical method of ensuring differential privacy is adding noise from a suitable kind of noise distribution to the data. Roughly speaking, the less privacy budget is provided, the higher the privacy remains but the more the utility suffers since this usually requires a higher amount of noise.
Formally, an algorithm $\mathcal{A}$ provides $\varepsilon$-differential privacy if for any two datasets $D_1$ and $D_2$ differing in at most the data of one individual, and for any subset of outputs $O\subseteq \text{Range}(\mathcal{A})$,  
	$Pr[\mathcal{A}(D_1) \in O] \,\leq\, \mathrm{e}^{\varepsilon} \cdot 
	Pr[\mathcal{A}(D_2) \in O]$.

Differential privacy was originally designed to provide privacy of aggregations, e.g., query counts or histograms, by adding noise to the counts in a differentially private manner. To make use of differential privacy for synthetic data generation, most of the models obfuscate the learned (latent) distributions by applying differentially private noise; synthetic trajectories are then generated by sampling from obfuscated distributions \cite{he_dpt_2015-1, roy_practical_2016, gursoy_utility-aware_2018-1, gursoy_differentially_2019-1, gursoy2020utility, liu_adaptive_2019, ghane_tgm_2020, deldar_enhancing_2020, wang_daily_2022, sun_synthesizing_2023, zhao_synthesizing_2019, mir2013dp, chen2012differentially, chen_differentially_2012-3, li2016differentially, chen2020rnn, sakuma2021trajectory}. 
Recently, models with \textit{local} differential privacy have been proposed  \cite{yang_2022, du_ldptrace_2023}, where data is perturbed locally, typically on users' devices, 
and the \textit{untrusted} collector and aggregator can only access the noisy records.

The authors of \cite{lestyan_search_2022} and \cite{yu_reconciling_2019} both propose neural networks that make use of differentially private stochastic gradient descent \cite{abadi_deep_2016}. In every model update, the gradients of all model parameters are clipped and Gaussian noise is added to the clipped gradients. 
Alatrista et al. \cite{10.1007/978-3-031-13448-7_7} use the architecture of DP-GAN \cite{HO2021103066} which is a novel approach to introduce DP into GAN architectures by extending the classical GAN setting to a three-player game with an additional classifier that aims to label data with respect to differential privacy constraints.
\cite{blanco-justicia_generation_2022} also consider privacy in their LSTM architecture without any privacy guarantees or privacy evaluations.
They choose one of the top-\textit{k} predictions uniformly at random for next-point generation to mitigate the learning of a 1-to-1 representation of the raw data. 

The original definition of differential privacy is based on the assumption that every user contributes only one record to a dataset as, for example, in a customer database. This does not necessarily hold for mobility data as a user can usually contribute an arbitrary number of trajectories, especially if waypoint trajectories are considered. This is a noteworthy issue, as differential privacy mechanisms based on the above assumption would only protect the privacy of a single trajectory (item-level privacy), but not the privacy of a user (user-level privacy) as formulated in the definition given above. 
Most coded publications that utilize waypoint trajectories do not consider this issue (e.g., \cite{ghane_tgm_2020}). Lesty{\'a}n et al. \cite{lestyan_search_2022} at least point out that their approach only provides item-level differential privacy and only Yu \cite{yu_reconciling_2019} provides a solution to guarantee user-level privacy. During the training phase of the neural network, they adjust differentially private stochastic gradient descent such that not a single record, i.e., trajectory, is used for training; instead, for each sampled user the gradient of the model parameters is obtained for a batch of the user’s data.
The second publication that carefully considers user-level privacy is by Mir et al. \cite{mir2013dp}. They provide differential privacy by applying controlled noise to a set of defined probability distributions, such as the distribution of home locations, and probabilities of a record at each location per hour. For each probability distribution, they evaluate the potential maximum user contribution and apply noise accordingly to satisfy user-level privacy.
All other publications only provide user-level privacy assuming that each user only contributes a single trajectory, which can potentially hold depending on the dataset, likely more so for staypoint trajectories. When models are applied in practice, the structure of the utilized dataset should always be revised in that respect.

Bindschaedler and Shokri \cite{bindschaedler_synthesizing_2016-1} provide \textit{plausible deniability} 
\cite{bindschaedler2017plausible} which ensures that there are $k$ alternative trajectories that could 
have produced a similar synthetic trajectory generated from a seed. This notion is based on the condition that each synthetic trajectory originates from a specific seed trajectory within the real data. As stated by the authors, this privacy guarantee protects against location inference attacks \cite{shokri_quantifying_2011} and membership inclusion attacks (which learns whether a particular 
individual with certain semantic habits has been in the seed dataset).

\textbf{Privacy evaluations.} Privacy evaluations use attack scenarios such as location inference attacks mentioned above \cite{bindschaedler_synthesizing_2016-1,zhao_synthesizing_2019} to estimate the success of a potential  
attack with and without privacy enhancement. The trajectory-user linking task identifies users from 
trajectories and links trajectories 
to them, thus a decrease in the accuracy of a state-of-the-art trajectory-user linking algorithm can be interpreted as an increase in privacy \cite{rao_lstm-trajgan_2020-1}. Zhan et al. \cite{zhan2022privacy} evaluate re-identification 
inaccuracy and Zhao et al. \cite{zhao_synthesizing_2019} social relationship based de-anonymization attacks. Du et al. \cite{du_ldptrace_2023} investigate their approach with regard to re-identification attack and the outlier attack. 

Another approach compares the similarity between real and synthetic traces assuming that a high similarity resembles low privacy. Different methods are used for this comparison: (1) measuring the distance between any real to any synthetic trajectory 
\cite{berke_generating_2022}, (2) quantifying how many trajectories in the real dataset exactly match trajectories 
in the synthetic dataset \cite{smolak_population_2020}, and (3) measuring the mutual information 
\cite{chen2020rnn, li2016differentially}. The sensitive locations disclosure attack is designed by 
\cite{deldar_enhancing_2020} which indicates how much the sensitive locations (points) of an original 
trajectory are similar to those of synthetic trajectories. 

Kulkarni et al. \cite{kulkarni_generative_2018-1} evaluate privacy with a location-sequence attack which shows to what level of accuracy trajectories can be reconstructed, and a membership inference attack which evaluates the adversary’s accuracy of inferring whether a target individual contributed to a 
dataset.

\subsection{Utility evaluation measures}
\label{sec:similarity}

There are mainly two different approaches to evaluate the utility of a synthetic data generation algorithm: (1) \textit{downstream tasks} and (2) \textit{(dis)similarity} of raw and synthetic data. The utility is considered high if the synthetic dataset is similar to the original dataset or performs similarly well on the downstream task.

Downstream tasks are a common method for utility evaluation of synthetic data in other areas, such as fraud detection \cite{finance_synthetic} in the financial sector or diagnostics~\cite{benaim_analyzing_2020, chen_synthetic_2021} for health data. Mobility data, however, is rarely used for similar prediction tasks, making it less straightforward to evaluate on downstream tasks. Nevertheless, a few of the coded papers include such evaluations based on the following tasks: \textit{road map updating} \cite{cao2021generating}, a task that aims to discover uncharted road segments in digital maps, \textit{next-location prediction} \cite{kulkarni_generating_2017-1, wei2020we, zhan2022privacy, jiang_continuous_2023}, a \textit{COVID-19 spreading simulation} \cite{feng_learning_2020, xusimulating, xiong_trajsgan_2023} and \textit{traffic control simulation} \cite{jiang_continuous_2023}, evaluating how a one-day construction site would change the mobility in that area.

\begin{figure}
     \Description[Categorization of utility evaluation measures in the coded literature]{<The utility evaluation measures can be structured in a tree representation: They first divide into `Downstream task' and `Similarity measures'. Similarity measures can further be sub-categorized into `Per-instance similarity', `Visual comparison', and `Statistical similarity'.>} 
    \centering
     \includegraphics[width=10cm]{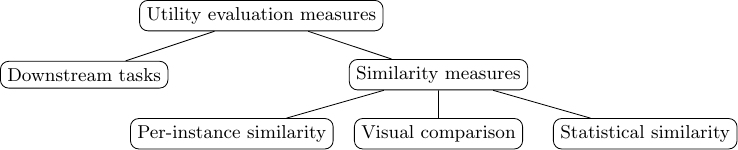}
	\caption{Categorization of utility evaluation measures in the coded literature.}
	\label{fig:tree}
\end{figure}

The similarity of raw and synthetic mobility data is more commonly evaluated. While there are standard (dis)similarity measures that have been proven to perform close to a human evaluation of synthetic data like synthetic images, the evaluation of synthetic mobility data is rather difficult and heterogeneous in the field \cite{benarous_synthesis_2022-1}. To structure the various approaches, we employ the categorization of  \cite{benarous_synthesis_2022-1} who consider the following types of evaluation measures for utility:
(1) \textit{per-instance similarity}, 
(2) \textit{visual comparison},
and (3) \textit{statistical similarity}  
(see Figure \ref{fig:tree}).

Per-instance similarity measures require the possibility to link a synthetic trajectory to its original counterpart to compare the two trajectories directly. Different authors make use of the Hausdorff distance \cite{rao_lstm-trajgan_2020-1, chen2020rnn, li2016differentially, jiang_continuous_2023}, the average distance between point pairs \cite{pang_development_2020, inproceedings, wei2020we}, or the BLEU (bilingual evaluation understudy) score and its successor the METEOR (Metric for Evaluation of Translation with Explicit ORdering) score, both originally designed to evaluate natural language machine translation  \cite{choi_trajgail_2021}. The Jaccard index is used to compute the similarity of activity spaces (i.e., areas individuals move within in the course of the day) between two trajectories \cite{rao_lstm-trajgan_2020-1, pang_development_2020}, and dynamic time warping is used to evaluate the reconstruction accuracy between two trajectories \cite{sakuma2021trajectory}.
However, most synthesis algorithms are not created such that a synthetic trajectory has a unique raw trajectory counterpart, thus per-instance similarity measures cannot be applied. Moreover, for privacy reasons, it is usually not an aspired goal to closely reproduce original trajectories\footnote{Also note that a high per-instance similarity does not assure that all original trajectories are `mocked up' in the synthetic data, which is why  \cite{benarous_synthesis_2022-1} consider an additional type of measures for \textit{diversity}. However, none of the coded papers considered those.}. From an application perspective, the utility should be considered high if distributions of relevant mobility characteristics of the dataset remain intact. 

Visual comparisons of graphs provide a straightforward approach to comparing such distributions, as done in the coded literature by looking at histograms displaying the number of visits \cite{wang_large_2021, zhao_synthesizing_2019, kulkarni_generative_2018-1}, length or speed of trajectories \cite{li_fts_2016, chiesa_2022}, or a selection of trajectories that indicates how well they match the road network \cite{kulkarni_generating_2017-1, wang_large_2021}. While such graphs and maps provide a great intuition about distributions of different mobility characteristics and deviations within the synthetic dataset, they are not suited to compare multiple synthesis approaches precisely and objectively.

Statistical (dis)similarity measures aim to provide single numbers that capture the similarity or dissimilarity of a certain characteristic between two datasets. For brevity, we include both under the term similarity measure. The mobility characteristics that were evaluated in the coded literature are presented in the next Section \ref{mobilitycharacteristics} and the statistical similarity measures used to compare them will be discussed in the following Section \ref{statisticalsimilarity}.

\subsubsection{Mobility characteristics}
\label{mobilitycharacteristics}
\begin{figure}[h]
     \Description[Taxonomy of statistical similarity measures.]{<The taxonomy of statistical similarity measures is defined by us in the following tree structure: The root consists of a `Mobility characteristic', e.g., `Spatial distribution'. One level lower, this is operationalized, e.g., by the `location popularity ranking', or `visits per location'. Each operationalization can be measures with concrete similarity measures, e.g., the location popularity ranking with the `Kendall-tau coefficient', or the `visits per location' with `Kullback-Leibler divergence', or the `Jensen-Shannon divergence'.>} 
    \centering
     \includegraphics[width=12cm]{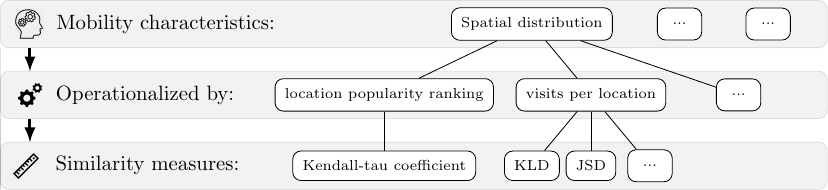}
	\caption{Taxonomy of mobility characteristics and statistical similarity measures.}
	\label{fig:taxonomy}
\end{figure}

We categorize the different mobility characteristics as 
\textit{spatial distribution}, \textit{trip lengths}, \textit{temporal distribution}, \textit{spatio-temporal distribution}, \textit{OD flows}, \textit{travel patterns} and \textit{user patterns} (see first column in Table \ref{table:similarity}). Each of these mobility characteristics can be operationalized by a specific feature that represents the mobility characteristic. See Figure \ref{fig:taxonomy} for an explanation of the taxonomy used here.

The \textit{spatial distribution} is the most commonly evaluated characteristic. It describes the distribution of locations (unrelated to a temporal component) and 
can be operationalized either by a  random selection of locations (called query error), the number of location visits, or the ranking of locations.

\textit{Trip lengths} are mostly operationalized as the distance with each trip which can either refer to the straight-line distance from the origin to the destination or the summed length of the distances between consecutive waypoints. 
Alternatively, the sequence length (i.e., number of points) or the trajectory diameter are considered, where the diameter is defined by any two (not necessarily consecutive) most distant points within a trajectory. Blanco-Justicia et al. \cite{blanco-justicia_generation_2022} further consider the interesting aspects of the average and maximum distance between consecutive points, as well as the distinct locations per trajectory, though they do not compute similarity measures for these statistics. 

As many approaches discard timestamps, the evaluation of \textit{temporal distributions} is only seen rarely, usually by comparing the number of trips per hour of day or the stay duration. 
The interlink between \textit{spatial and temporal distributions} is mostly analyzed by comparing the visits per location and hour of day or the stay duration per location.

\textit{OD flows} are a vital information for many urban mobility analyses, though only eight of the coded papers include such a similarity measure. 
By \textit{travel patterns} we refer to the order of locations in a dataset. It is typically 
operationalized through analyses of empirical distributions of  the most frequent subsequences of a given size, e.g., the top 20 subsequences of length 3, or by comparing the rankings. 
\textit{User mobility patterns} describe mobility behavior of individuals. They are captured by the comparison of the user activity space (e.g., radius of gyration), number of distinct locations per user (and day), I-Rank (i.e., the frequency of visiting the personal top one or more locations), number of trips per day and user, spatio-temporal visitation patterns per user and hour of day, the mobility entropy (which informs about the degree of predictability of an individual’s whereabouts; it characterizes the heterogeneity of individuals' visitation patterns), or the semantic similarity of user movements according to temporal patterns.

In addition to the presented similarity measures that can be applied to compare any two mobility datasets, some measures are only suited for specific approaches because additional information is assumed or needs to be  inferred, such as point of interest (POI) categories \cite{rao_lstm-trajgan_2020-1},  gender \cite{roy_practical_2016}, subscription payment plans \cite{roy_practical_2016}, closest matched road segment \cite{cao2021generating}, proportion of ordinary and express ways \cite{cao2021generating}, inferred home and work locations  \cite{berke_generating_2022}, proportion of traffic violations \cite{sun_synthesizing_2023},  cluster \cite{dandekar_trajectory_2016}, and  inferred friendship between users \cite{zhao_synthesizing_2019}. 

\subsubsection{Statistical similarity}
\label{statisticalsimilarity}

A variety of measures is utilized in the coded works to capture statistical similarity between operationalizations of mobility characteristics in real and synthetic data.

The \textit{Kullback-Leibler divergence (KLD)} \cite{kullback_information_1951}, also called relative entropy, is a widely used statistic to measure how far 
a probability distribution $P$ deviates from a reference probability distribution $Q$ on the same probability space $\mathcal{X}$.
For discrete distributions $P$ and $Q$, it is formally defined as 
$$D_{KL}(P||Q):= \sum\limits_{x \in \mathcal{X}: P(x)>0} P(x)\cdot \log \frac{P(x)}{Q(x)}.$$

For example, the spatial distribution can be evaluated by comparing the share of records in the synthetic data ($P$) per tile ($x$) of the tessellation ($\mathcal{X}$) to the share of records computed on the real data ($Q$). 
The larger the deviation of $P$ from $Q$, the larger the value of the resulting KLD, with a minimum value of $0$ for identical distributions.

Note that KLD is not symmetric, i.e., $D_{KL}(P||Q)~\neq~D_{KL}(Q||P)$, which is why KLD is best applicable in settings with a reference model $Q$ and a fitted model $P$. However, the lack of symmetry implies that it is not a distance metric in the  mathematical sense.  

The \textit{Jensen-Shannon divergence (JSD)} solves this issue and builds on the KLD to calculate a symmetrical score.
Additionally, JSD provides a smoothed and normalized version of KLD, with scores between 0 (identical) and 1 (maximally different) when using the base-2 logarithm, thus making it easier to relate the resulting score within a fixed finite range. 

It is worth noting that KLD is only defined if $Q(x)\neq 0$ for all $x$ in the support of $P$, while this constraint is not required for JSD.
In practice, both KLD and JSD are computed for discrete approximations of continuous distributions, e.g., histograms approximating the number of trips over time based on daily or hourly counts. However, the choice of histogram bins has an impact in two respects:
Say we want to compare the number of visits per tile. Depending on the granularity of the chosen tessellation, there might be tiles with $0$ visits in the real dataset but $>0$ visits in the synthetic dataset, thus KLD would not be defined for such cases.
Additionally, the resulting values for both KLD and JSD vary according to the choice of bins, e.g., by reducing the granularity of the tessellation, the values of KLD and JSD will tend to be smaller. 

Both KLD and JSD do not account for a distance of instances in the probability 
space $\mathcal{X}$. Consider the spatial distribution of a mobility dataset that differs from another 
dataset only within the values of two tiles. In one case these two tiles are direct neighbors, in a second case they are far apart. Intuitively, the two distributions that differ in neighboring tiles are more similar, 
but JSD and KLD yield the same value in both cases.
In contrast, the \textit{earth mover's distance (EMD)} between two empirical distributions  
allows to take the underlying geometry of the space into account. 
The EMD is proportional to the minimum amount of work required to convert one distribution into the other \cite{levina_earth_2001}. 
The amount of work is determined by the defined distance between instances (e.g., tiles or histogram bins), thus, it allows for an intuitive interpretation.
For example, if the EMD of two spatial distributions is computed based on the geographic straight-line distance between tile centers in meters, an EMD of 100 signifies that on average each record of the first distribution needs to be moved 100 meters to reproduce the second distribution. On the downside, there is no fixed range as  for the JSD which provides values between 0 and 1. Thus the EMD always needs to be interpreted in the context of the dataset and the EMD of different datasets cannot be compared directly.

The \textit{mean squared error (MSE)} is a common error measure, computing  the average squared difference between  observed (i.e., real) and predicted (i.e., synthetic) values. Unlike KLD and JSD, it is not explicitly designed for probability distributions. 
The number of averaged values can correspond to single records, e.g., comparing the real observed trip length of a trajectory to its synthetic counterpart, or to histogram bins, e.g., grouping trip lengths into $5km$ bins and comparing the count of each corresponding bin. Apparently, the range of the MSE highly varies based on the implementation choice.
The reviewed literature also uses variations of the MSE such as the \textit{root MSE (RMSE)} and \textit{standard root MSE (SRMSE)}.

The \textit{cosine similarity} is also used, but less commonly. The general idea is to consider two vectors as similar, if their angle is small, thus both pointing in a similar direction,  while two orthogonal vectors have zero similarity, and the corresponding distributions would be considered unrelated. It is frequently applied in text analysis to, e.g., measure document similarity, having the advantage of working well even with very sparse data such as document representations based on word counts. The tendency of mobility data to develop sparse distributions as well makes cosine similarity a good candidate also for this context \cite{han_2_2012}. 

The \textit{Kendall's $\tau$ coefficient}, also known as Kendall rank correlation coefficient, 
is a measure of the strength and direction of association that exists between two variables measured on an ordinal scale.
It returns a value between 0 and 1, where 0 means no relationship and 1 is a perfect relationship, determining the strength of association based on the pattern of concordance (ordered in the same way) and discordance (ordered differently) between all pairs.
In the case of mobility data, it can, for example, be used to measure discrepancies in locations’ popularity ranking \cite{gursoy_utility-aware_2018-1}, where the popularity of a location $L$ is defined as the number of times $L$ is visited by trajectories in the considered dataset. 
Some approaches also utilize the \textit{Pearson correlation} to compare two histograms, where each histogram bin corresponds to one `sample point', 
and the correlation is measured between the counts falling into each respective bin based on the real dataset and those based on the synthetic one.

\renewcommand*{\arraystretch}{1.2}
\begin{footnotesize}
\begin{longtable}{p{13mm}p{47mm}p{68mm}}
		\caption{Statistical similarity measures used by the coded publications and categorized by mobility characteristics. Sometimes mobility characteristics are operationalized similarly but named differently, thus different name variations are listed.\label{table:similarity}} \\
\hline 
\textbf{Mobility char.} & \textbf{Operationalized by} & \textbf{Similarity measures}
\\ \hline
spatial \newline distribution & query error (= count error = count query) &
  relative error for spatial counting queries, averaged over $n$ queries of randomly selected locations \cite{gursoy_utility-aware_2018-1, gursoy_differentially_2019-1, chen2012differentially, chen_differentially_2012-3, bindschaedler_synthesizing_2016-1, deldar_enhancing_2020, liu_adaptive_2019, du_ldptrace_2023, sun_synthesizing_2023, yang_2022} \\
 & visits per location (= road segment usage = density error) &
  KLD \cite{bindschaedler_synthesizing_2016-1,pappalardo_data-driven_2018-1, zhao_synthesizing_2019}/JSD \cite{ouyang_non-parametric_2018-1, xusimulating, wang_large_2021, feng_learning_2020, du_ldptrace_2023, jiang_continuous_2023, xiong_trajsgan_2023}/EMD \cite{mir2013dp}/RMSE \cite{tamura_synthetic_2022, bwambale_getting_2021, pappalardo_data-driven_2018-1}/cosine similarity \cite{cao2021generating}/SRMSE \cite{badu-marf_composite_2020}/ARE \cite{wang_privtrace_2022}\\

 & top $n$ coverage of locations &
    \% of common locations in top $n$ loc.
  \cite{bindschaedler_synthesizing_2016-1}/hotspot error \cite{du_ldptrace_2023} \\
 &
  location popularity ranking &
  Kendall's $\tau$ coefficient \cite{gursoy_utility-aware_2018-1, deldar_enhancing_2020, du_ldptrace_2023} \\ 
   &
  spatial semantic information &
  recall of top-\textit{k} visits overlap of travel purposes \cite{xiong_trajsgan_2023} \\ \hline

trip lengths &

  trip length (= distance = length error)  &
  JSD \cite{gursoy_utility-aware_2018-1, feng_learning_2020, xusimulating, wang_large_2021, deldar_enhancing_2020, liu_adaptive_2019, sun_synthesizing_2023, wang_privtrace_2022, jiang_continuous_2023, xiong_trajsgan_2023, yang_2022}/KLD \cite{berke_generating_2022, smolak_population_2020, pappalardo_data-driven_2018-1, kang_trag_2021, du_ldptrace_2023}/RMSE \cite{pappalardo_data-driven_2018-1}/EMD \cite{anda_synthesising_2021}/Pearson correlation \cite{badu-marf_composite_2020} 
  \\
 & sequence length (number of points) &
  JSD 
  \cite{wang_large_2021, deldar_enhancing_2020, lestyan_search_2022} \\
 & diameter error &
  JSD \cite{gursoy_utility-aware_2018-1, gursoy_differentially_2019-1, he_dpt_2015-1, ghane_tgm_2020, deldar_enhancing_2020, liu_adaptive_2019, yu_reconciling_2019, du_ldptrace_2023, sun_synthesizing_2023, wang_privtrace_2022} \\ \hline

temporal &

  trips per hour (= activity start time) & 
 KLD \cite{pappalardo_data-driven_2018-1}/RMSE \cite{pappalardo_data-driven_2018-1}/Pearson correlation \cite{rao_lstm-trajgan_2020-1, tamura_synthetic_2022}/EMD \cite{anda_synthesising_2021}\\

distribution & 
stay duration &  KLD \cite{pappalardo_data-driven_2018-1, kang_trag_2021}/RMSE \cite{pappalardo_data-driven_2018-1}/JSD \cite{ghane_tgm_2020}/EMD \cite{anda_synthesising_2021}\\ \hline

spatio-temporal &

  spatio-temporal distribution of location visits (= spatial error [per hour of day]) & JSD \cite{ouyang_non-parametric_2018-1, xusimulating, xiong_trajsgan_2023}/EMD \cite{smolak_population_2020, lestyan_search_2022, anda_synthesising_2021}/cosine similarity \cite{cao2021generating}/correlation of visits by hour of day \cite{rao_lstm-trajgan_2020-1, tamura_synthetic_2022, pang_development_2020}\\
  
 distribution & spatio-temporal mobility features &
  aggregate mobility model \cite{bindschaedler_synthesizing_2016-1} \\
 & stay duration per location &
  JSD \cite{feng_learning_2020, ouyang_non-parametric_2018-1, xusimulating}/Pearson correlation \cite{berke_generating_2022}/KLD \cite{bindschaedler_synthesizing_2016-1} \cite{berke_generating_2022}\\ \hline

OD flows &

  trips per OD pair (= trip error) &
  JSD \cite{gursoy_utility-aware_2018-1, gursoy_differentially_2019-1, he_dpt_2015-1, deldar_enhancing_2020, yu_reconciling_2019, du_ldptrace_2023, sun_synthesizing_2023, jiang_continuous_2023}/EMD \cite{lestyan_search_2022}/cosine similarity \cite{cao2021generating}\\
 & realistic OD pairs &
  \% of OD pairs non existent in the original data \cite{berke_generating_2022} \\
 & trips per OD pair for each hour of day &
  EMD  \cite{lestyan_search_2022} \\ \hline
  
travel \hspace{1cm} patterns &

  counts per travel pattern (= freq. patterns) &
  average relative error \cite{gursoy_utility-aware_2018-1, gursoy_differentially_2019-1, chen_differentially_2012-3, sun_synthesizing_2023}/EMD \cite{anda_synthesising_2021}/JSD \cite{choi_trajgail_2021} \\
 & top $k$ most frequent patterns & set similarity between top $k$ patterns (F1 Score) \cite{gursoy_utility-aware_2018-1, he_dpt_2015-1, ghane_tgm_2020, yu_reconciling_2019, du_ldptrace_2023, yang_2022}/true positive ratio
 \cite{lestyan_search_2022}/ARE \cite{wang_privtrace_2022} \\
 & frequent pattern popularity ranking &
  Kendall's $\tau$ coefficient \cite{gursoy_differentially_2019-1, deldar_enhancing_2020} \\
 & route distribution &
  EMD of inner cells (waypoints) for each OD pair \cite{lestyan_search_2022}\\ \hline
  
user  &
  activity space (e.g., radius of gyration) & 
  JSD \cite{feng_learning_2020, xusimulating, jiang_continuous_2023, xiong_trajsgan_2023}/KLD \cite{pappalardo_data-driven_2018-1}/RMSE \cite{pappalardo_data-driven_2018-1}/EMD \\
patterns & number of distinct locations per user [and day] (= DailyLoc) &
  JSD \cite{feng_learning_2020, xusimulating}/KLD \cite{berke_generating_2022, pappalardo_data-driven_2018-1}/RMSE \cite{pappalardo_data-driven_2018-1} \\
  & user location frequency (= I-Rank) &
  JSD \cite{feng_learning_2020}/KLD \cite{pappalardo_data-driven_2018-1}/ RMSE \cite{pappalardo_data-driven_2018-1} \\
 & number of trips per day \& user &
  KLD \cite{pappalardo_data-driven_2018-1}/RMSE \cite{pappalardo_data-driven_2018-1}/EMD \cite{anda_synthesising_2021} \\
 & stay duration per location \& user &
  JSD \cite{zhou_toward_2021-1} \\
 & spatio-temporal pattern per user &
  JSD \cite{zhou_toward_2021-1} \\
 & mobility entropy &
  EMD \cite{anda_synthesising_2021}/KLD \cite{pappalardo_data-driven_2018-1}/RMSE \cite{pappalardo_data-driven_2018-1}\\
 & semantic similarity &
  EMD of semantic cluster membership distribution \cite{anda_synthesising_2021}\\  
\end{longtable}
\end{footnotesize}

\section{Assessment and comparison} 
\label{sec:assessment}
As described in Section \ref{sec:datasets},
mobility data can take various forms and represent different aspects of human mobility. Most of the coded works formulate, mostly rather implicitly, assumptions regarding trajectory semantics or additional knowledge
typically required to implement a certain algorithmic idea or to achieve solid performance for a certain type of use case. These additional constraints, however, are also a major criterion for the choice of method in practice and additionally yield limitations to the level of comparability of existing approaches. We identified the `granularity level' of mobility as a suitable criterion to compare the coded methods within the respective groups from an application perspective, supporting practitioners to identify the most suitable approach for their use case. More specifically, in an iterative process we defined the following three categories and assigned the synthesis methods to one of them based on careful reading and inspection of motivational examples, used datasets, and model assumptions: 
 (1) \textit{trips}, (2) \textit{user movements}, and (3) \textit{city population}. 
We find these categories to provide a suitable starting point to narrow down the selection of potential methods, given a dataset and an aspired application\footnote{Even though the used dataset provides a good indicator for the category, this is not the decisive factor. For example, a continuously sampled GPS trace can either be cut into multiple waypoint trips or a single staypoint trajectory could be extracted that resembles the user movement. User movement and city population could potentially make use of similar datasets but is mainly categorized based on the underlying motivation.}.
 Some approaches are too generic or do not provide enough information to group them according to one of our categories, and have thus been classified as `unspecified'. Table \ref{table:overview} presents all coded publications grouped by these categories. 
 
A taxi trip comprises the typical understanding of single-trip mobility: a trajectory lasting a couple of minutes where an origin and destination location are connected with a route following the street network. On the contrary, a trajectory in terms of user movements is comprehended as a sequence of semantically meaningful stay locations over one or multiple days, for example, recorded by social network check-in data like the Foursquare dataset. The city population category utilizes trajectories to create representative mobility data for a (group of the) citywide population, as for example queried in a survey. The original trajectory data is thereby not necessarily seen as the ground truth but only as a (biased) piece of information next to other data sources used for a realistic mobility representation. 
In addition to the categorization, this chapter provides a basic understanding of each method and, if possible, compares them to further facilitate the choice of the algorithm within a category.

\begin{figure}[ht]
     \Description[Timeline of all coded publications]{<Timeline of all coded publications, including the information about how often a publication was cited, an arrow to all publications used for benchmark, and an indication whether the model is deep learning based. The first publication, ngram, from 2012 is also the one with most citations, which are more than 200. It is used as benchmark by AdaTrace, DeepSyntheziser and DP-Loc. From 2013 to 2018 a series of well known publications with privacy guarantees followed, namely, DP-WHERE, DPT, SGLT and AdaTrace. From 2019 a large set of deep learning approaches without privacy guarantees were published.>} 
    \centering
     \includegraphics[width=14cm]{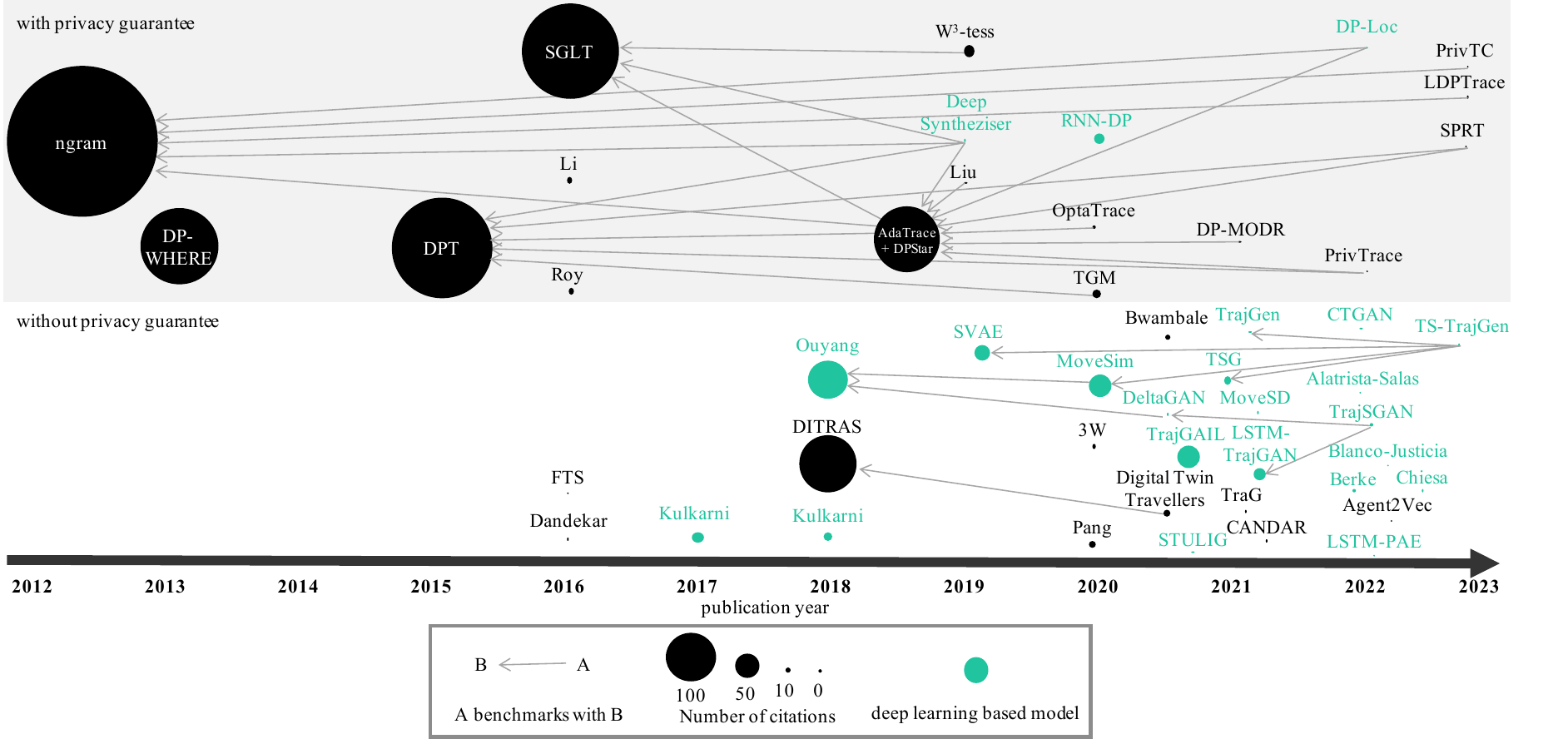}
    \caption{Timeline of all coded publications, displaying the model's name (if existent) or the name of the first author. The size of the bubble indicates how often the publication was cited (based on Google Scholar, accessed 06.03.2023). Models providing privacy guarantees and use of deep learning algorithms are indicated, also, arrows indicate when a previous model was used as a benchmark.}
    \label{fig:timeline}
\end{figure}

Figure \ref{fig:timeline} presents an overview of the coded works, sorted by the publication year and indicating the number of citations.
The works of Chen et al. \cite{chen2012differentially, chen_differentially_2012-3} are (one of) the first and most widely cited approaches (308 and 297 citations according to Google Scholar, 06.03.2023) in this field that have been used as benchmarks by multiple authors \cite{gursoy_utility-aware_2018-1, lestyan_search_2022, deldar_enhancing_2020}. As many further publications build on this rather general method, we start by introducing the main idea.

The authors initially \cite{chen2012differentially} proposed a prefix tree structure to store the counts of each sequence of a given length, with noise added to ensure differential privacy. From the noisy prefix tree synthetic trajectories are reconstructed. This has been later  extended \cite{chen_differentially_2012-3} to an approach commonly referred to as \textit{Ngram} by allowing  
variable-length $n$-grams; their main purpose is to prevent quickly decreasing utility (due to the implemented DP mechanism) when the noisy prefix tree grows by pruning the prefix tree.  The algorithm is potentially applicable to any type of trajectory, though it does not scale well with long trajectories and many different unique locations, thus being better suited for shorter staypoint trajectories (like public transport smart card data used for evaluation in the respective publication).

%\pagebreak
\renewcommand*{\arraystretch}{1}
\begin{footnotesize}
	\begin{longtable}[t]{p{1mm}p{30mm}p{20mm}p{6mm}p{13mm}p{15mm}p{29mm}}
		\caption{Overview of coded publications, grouped by category, including the model name, publication year, privacy considerations, representation of time (`?' signifies that no clear information is provided), and applied algorithms. * Source code is available and respective links are provided in the supplemental materials.	\label{table:overview}} \\	
		\hline
		\centering
		& \textbf{Name} & \textbf{Model name} & \textbf{Year} & \textbf{Privacy} & \textbf{Time} & \textbf{Algorithm}  \\
		\hline
		\multirow{25}{=}{\rotatebox[origin=c]{90}{\textbf{Trips}}}
		& He et al. \cite{he_dpt_2015-1} &DPT &
		2015 &
		DP &
		- &Markov model\\
		&Roy et al. \cite{roy_practical_2016} &
		- &
		2016 &
		DP &
		- &
		prob. dist.
		\\
		&
		Li et al. \cite{li_fts_2016} &
		FTS &
		2016 &
		- &
		? &
		heuristics
		\\
		&
		Gursoy et al. \cite{gursoy_utility-aware_2018-1, gursoy_differentially_2019-1} &
		AdaTrace/ \newline DP-Star* &
		2018/ 2019 &
		DP, priv. eval. &
		- &
		Markov model, prob. dist.
		
		\\
		
		&
		Yu \cite{yu_reconciling_2019} &
		Deep Syntheziser &
		2019 &
		DP &
		- &
		RNN
		\\
		&
		Liu et al. \cite{liu_adaptive_2019} &
		- &
		2019 &
		DP &
		- &
		Markov model
		\\
		&
		Huang et al. \cite{inproceedings} & SVAE & 2019 & - & - & VAE, LSTM
		\\
		&
		Gursoy et al. \cite{gursoy2020utility} &
		OptaTrace &
		2020 &
		DP &
		- &
		Markov model, prob. dist. 
		\\
		&
		Ghane et al. \cite{ghane_tgm_2020} &
		TGM &
		2020 &
		DP &
		- &
		Markov model, prob. dist.
		\\
		
		&
		Kang et al. \cite{kang_trag_2021} &
		TraG &
		2021 &
		- &
		- &
		prob. dist.
		\\
		&
		Wei et al. \cite{wei2020we} &
		MoveSD &
		2021 &
		- &
		? &
		GAIL (GAN + Inverse RL)
		\\
		&
		Choi et al. \cite{choi_trajgail_2021} &
		TrajGAIL* &
		2021 &
		- &
		- &
		GAIL (GAN + Inverse RL) 
		\\
		&
		Wang et al. \cite{wang_large_2021} &
		TSG* &
		2021 &
		- &
		- &
		GAN, enc.-dec., CNN, LSTM
		\\
		
		&
		Cao \& Li \cite{cao2021generating} &
		TrajGen &
		2021 &
		- &
		start time &
		enc.-dec., DCGAN, ANN
		\\
		&
		Deldar \& Abadi \cite{deldar_enhancing_2020} &
		DP-MODR(T) &
		2021 &
		DP &
		start time &
		prob. dist. 
		\\
		&
		Lestyán \cite{lestyan_search_2022} &
		DP-Loc* &
		2022 &
		DP &
		start time &
		VAE, FFN
		\\
		&
		Badu-Marfo et al. \cite{badu-marf_composite_2020} &
		CTGAN &
		2022 &
		- &
		- &
		GAN, RNN
		\\
		&
		Du et al. \cite{du_ldptrace_2023} &
		LDPTrace* &
		2023 &
		local DP, priv. eval. &
		- &
		Markov model, prob. dist.
		\\
		& 
		Wang et al. \cite{wang_privtrace_2022} &
		PrivTrace* &
		2022 &
		DP &
		- &
		Markov model \\
		&
		Xiong et al. \cite{jiang_continuous_2023} &
		TrajSGAN &
		2022 &
		- &
		start time &
		GAN, seq2seq, CNN \\
		&
		Jiang et al. \cite{jiang_continuous_2023} &
		TS-TrajGen &
		2023 &
		- &
		start time &
		GAN \\
		&
		Sun et al. \cite{sun_synthesizing_2023}  &
		SPRT &
		2023 &
		DP &
		- &
		Markov model, prob. dist. 
		\\

		\hline
		
		\multirow{11}{=}{\rotatebox[origin=c]{90}{\textbf{User Movement}}}
		&Bindschaedler \& Shokri \cite{bindschaedler_synthesizing_2016-1} &
		SGLT &
		2016 &
		plausible deniability &
		time window &
		clustering, location semantic graph
		\\
		&
		Dandekar et al. \cite{dandekar_trajectory_2016} &
		- &
		2016 &
		- &
		- &
		clustering, random walk
		\\
		&
		Ouyang et al. \cite{ouyang_non-parametric_2018-1} &
		- &
		2018 &
		- &
		timestamp &
		GAN, CNN 
		\\
		&
		Zhao et al. \cite{zhao_synthesizing_2019} &
		W$^3$-tess &
		2019 &
		DP &
		time window &
		statistical similarity
		\\
		&
		Feng et al. \cite{feng_learning_2020} &
		MoveSim* &
		2020 &
		- &
		time window &
		GAN, CNN
		\\
		&
		Rao \cite{rao_lstm-trajgan_2020-1} &
		LSTM-TrajGAN* &
		2021 &
		priv. eval. &
		timestamp &
		GAN, RNN 
		\\
		&
		Xu et al. \cite{xusimulating} &
		DeltaGAN &
		2021 &
		- &
		timestamp &
		GAN, RNN 
		\\			&
		Zhou et al. \cite{zhou_toward_2021-1} &
		STULIG* &
		2021 &
		priv. eval. &
		? &
		VAE, CNN
		\\
		&
		Tamura et al. \cite{tamura_synthetic_2022} &
		Agent2Vec &
		2022 &
		- &
		time window &
		word embedd., clustering
		\\
		&
		Chiesa \& Taraglio \cite{chiesa_2022} &
		- &
		2022 &
		- &
		timestamp &
		VAE
		\\
		
		\hline

		\multirow{8}{=}{\rotatebox[origin=c]{90}{\textbf{City population}}}
		&
		Mir et al. \cite{mir2013dp} &
		DP-WHERE &
		2013 &
		DP &
		timestamp &
		prob. dist. 
		\\
		&
		Papparlado \& Simini \cite{pappalardo_data-driven_2018-1} &
		DITRAS* &
		2018 &
		- &
		time window &
		Markov model
		\\
		&
		Smolak et al. \cite{smolak_population_2020} &
		3W* &
		2020 &
		priv. eval. &
		timestamp &
		prob. dist.
		\\
		&
		Pang \cite{pang_development_2020} &
		- &
		2020 &
		- &
		time window &
		Inverse RL
		\\
		&
		Bwambale et al. \cite{bwambale_getting_2021} &
		- &
		2021 &
		- &
		? &
		prob. dist.
		\\
		&
		Anda et al. \cite{anda_synthesising_2021} &
		Digital Twin Travellers &
		2021 &
		- &
		timestamp &
		Markov model, prob. dist.
		\\
		
		&
		Berke et al. \cite{berke_generating_2022} &
		-* &
		2022 &
		priv. eval. &
		time window &
		RNN
		\\
		\hline
		
		\multirow{9}{=}{\rotatebox[origin=c]{90}{\textbf{Unspecified}}}
		&
		Chen et al.  \cite{chen2012differentially, chen_differentially_2012-3} &
		Ngram &
		2012 &
		DP &
		- &
		Markov model
		\\
		&
		Li et al.  \cite{li2016differentially} &
		- &
		2016 &
		DP &
		- &
		partition, clustering
		\\
		&
		Kulkarni \& Garbinato \cite{kulkarni_generating_2017-1} &
		-* &
		2017 &
		- &
		- &
		RNN 
		\\
		&
		Kulkarni et al. \cite{kulkarni_generative_2018-1} &
		- &
		2018 &
		- &
		- &
		RNN, GAN, copula 
		\\
		&
		Chen et al.  \cite{chen2020rnn} &
		RNN-DP &
		2020 &
		DP &
		- &
		RNN, partition, clustering
		\\
		&
		Sakuma et al. \cite{sakuma2021trajectory} &
		CANDAR &
		2021 &
		DP &
		timestamp &
		Autoencoder, LSTM
		\\
		&
		
		Zhan et al. \cite{zhan2022privacy} &
		LSTM-PAE &
		2022 &
		priv. eval. &
		? &
		Autoencoder, LSTM 
		\\
		&
		
		Blanco-Justicia et al. \cite{blanco-justicia_generation_2022} &
		- &
		2022 &
		top-\textit{k} \newline sampling  &
		- &
		LSTM 
		\\
		& 
		Alatrista-Salas et al. \cite{10.1007/978-3-031-13448-7_7} &
		- &
		2022 &
		DP &
		timestamp &
		GAN 
		\\
		& 
		Yang et al. \cite{yang_2022} &
		- &
		2022 &
		local DP &
		- &
		Hidden Markov model
		
	\end{longtable}
\end{footnotesize}

	\subsection{Trips}
	\label{uc:trips}
	
Models in this category consider mobility data as a collection of trips where each trip consists of a start and an end location that are connected with a route that (more or less) follows the road network. All approaches that require \textit{waypoints} as input fall into this category.

This (implicit) requirement can be deduced from various algorithmic designs: The models of \cite{ghane_tgm_2020, wei2020we, sun_synthesizing_2023} are based on the assumption that two consecutive locations stem from neighboring tiles;  \cite{cao2021generating, wang_large_2021, choi_trajgail_2021, du_ldptrace_2023} make use of a fixed sampling rate between all consecutive points; the approaches by \cite{gursoy_differentially_2019-1, kang_trag_2021, liu_adaptive_2019} are tailored to connect a previously selected origin and destination location, and  \cite{deldar_enhancing_2020, sun_synthesizing_2023} makes use of distributions of origins and trip lengths for trip generation; \cite{li2016differentially, he_dpt_2015-1, du_ldptrace_2023} rely on length or acceleration, and thus features that make use of waypoints.

Most approaches focus on reconstructing meaningful spatial sequences and dismiss the temporal dimension. If timestamps are generated, then usually only a start time, since successive locations follow in a short time interval when waypoints are assumed.
	
	\textbf{Two basic approaches use heuristics and probability distributions without providing any privacy
		guarantees:}	(1) \textit{FTS} (feature-based trajectory synthesis) \cite{li_fts_2016} cuts the trajectories into subtrajectories 
	and resamples them based on heuristics that utilize the computed length, speed, acceleration, 
	u-turn rate and density. 
	(2)  \textit{TraG} \cite{kang_trag_2021} generates  synthetic traces by drawing OD pairs from a distribution and then connecting the two 
	points with `context-aware waypoints'. They define waypoints as stop points from a 
	continuous trace where the moving object stops for longer than a given threshold, which they set to 60 seconds within their evaluation, aimed to detect waiting times due to traffic. They argue that urban 
	hotspots should correlate with bad traffic and thus 
	longer waiting times.
	With this approach they reduce the space of potential locations, only focusing on origins and destinations 	(based on a grid) and intense traffic nodes. If only such hotspots are of interest this might be a suitable 
	approach. If traffic flows are additionally of interest, this will likely not provide satisfying results. 

	\textbf{A set of models strives to create differentially private synthetic trajectories based on Markov 
		models and probability distributions:}
	Roy et al. \cite{roy_practical_2016} synthesize bike sharing data but consider only OD trips, meaning each trajectory consists of just two points, which is much simpler than synthesizing entire sequences.
	He et al. \cite{he_dpt_2015-1} present a system called \textit{DPT} (differentially private trajectories) based on the work of Chen et al. \cite{chen_differentially_2012-3} using a prefix structure. They argue that \cite{chen_differentially_2012-3} is only usable for a small domain of locations and introduce a novel hierarchical grid that adapts the resolution based on speed to optimize the number of counts maintained in the prefix tree.
	DPT is a highly recognized early work in this area with 207 citations (Google Scholar, 06.03.2023) and has 
	served as a popular benchmark \cite{gursoy_utility-aware_2018-1, gursoy_differentially_2019-1, ghane_tgm_2020, yu_reconciling_2019}.

Many model architectures follow a similar outline: a distribution of start locations is created, a Markov chain is used to sample single steps along the way and the end of the trajectory is determined, either by a conditional probability of destinations and/or trip lengths in relation to the start locations, or by including a termination condition in the transition probabilities. Differential privacy is guaranteed by applying noise to each distribution and to the transition probabilities. 
 Gursoy et al. presented \textit{AdaTrace} \cite{gursoy_utility-aware_2018-1} in 2018 and \textit{DP-Star} 
	\cite{gursoy_differentially_2019-1} in 2019, two very similar examples  of such an architecture.
 	Their approach has been seen as state-of-the-art and used as a benchmark by various authors
	\cite{lestyan_search_2022, liu_adaptive_2019, deldar_enhancing_2020, yu_reconciling_2019, sun_synthesizing_2023, wang_privtrace_2022}.
	In 2020, Gursoy et al. published \textit{OptaTrace} \cite{gursoy2020utility} which extends AdaTrace in 
	three ways: 
	(1) A Bayesian optimization module that iteratively searches for the optimized parameters that minimize 
	a given error (i.e., similarity measure), 
	(2) a utility module, that contains four categories of error metrics, 
	(3) a front-end web interface. 
 Liu \cite{liu_adaptive_2019} extended the Ngram approach with OD distributions and trajectory length distributions. Additionally, they focus on optimizing differential privacy budget on the hierarchical structure of the prefix tree using a sparse vector technique.
  \textit{TGM} \cite{ghane_tgm_2020}  (trajectory generative mechanism) represents trajectories which a graph, each node stores the prefix of the last \textit{k} stays and its stay time (understood in terms of how often the same node was also the consecutive node). Possible next steps are limited to the eight neighboring grid cells plus the current cell, considerably reducing the number of parameters of the Markov model. 
 \textit{DP-MODR} \cite{deldar_enhancing_2020} (differentially private mechanism for synthetic moving objects database release) includes an extension \textit{DP-MODRT} that allows the gerneration of 
time-dependent locations. To also include temporal information, DP-MODRT disaggregates start domain cells, their median lengths, and the cost matrix additionally according to start time windows.  \textit{LDPTrace} \cite{du_ldptrace_2023} is the first approach to implement local differential privacy. \textit{SPRT} \cite{sun_synthesizing_2023} (Synthesizing Private and Realistic Trajectories) includes the constraint of a geography-aware grid to only allow transitions that reassemble the road network. 
 \textit{PrivTrace} \cite{wang_privtrace_2022} employs first and second-order Markov chains. The authors claim to thereby reduce the disadvantages of AdaTrace which fails to obtain enough transition information with only a a first-order Markov chain model and DPT which introduces excessive noise due to DP with its high-order Markov chain model.

	As many of these models use one another as benchmarks, we can compare their performances to a certain degree:
	According to Gursoy et al., the authors of AdaTrace and DP-Star,  both approaches outperform DPT \cite{he_dpt_2015-1} and Ngram, especially in terms of similarity 
	measures capturing OD flows and the diameter of trajectories. The prefix tree approach of the latter two is not well suited to maintain OD relations; also, they find many loops and u-turns in their trajectories.
	Ghane et al. \cite{ghane_tgm_2020} criticize that DPT limits a trajectory to its first few points which 
	cannot represent the trajectory movement pattern properly. They find that their algorithm TGM outperforms DPT in maintaining travel distance and stay times. In terms of performance, they find DPT to be highly inefficient, as it consumes more than 13 GB of memory (vs 0.26 GB for TGM) and over 673h of runtime (vs. 233s for TGM) for  Porto Taxi. 
	To properly assess TGM, evaluations of
	the spatial distribution, OD flows, and trip lengths would be useful, which are lacking in its publication. Also, it would be interesting to see how it benchmarks against AdaTrace/DP-Star.
	
	\textit{DP-Loc} \cite{lestyan_search_2022} and DP-MODR  both use Ngram
    and AdaTrace as benchmarks and both find that Ngram outperforms 
	AdaTrace in terms of frequent patterns, while the authors of  AdaTrace, on the other hand, find their model to be superior. 
	In the context of frequent patterns, however, it should be noted that both Ngram and AdaTrace tend to create rather short trajectories \cite{lestyan_search_2022}.
	The results for the similarity of the spatial distribution are also inconsistent: while the authors of AdaTrace state that it outperforms Ngram in terms of query error and location ranking, the DP-Loc paper shows a clear superiority of Ngram based on the EMD, and Deldar and Abadi (DP-MODR) find that it depends on $\epsilon$ and the grid resolution which approach outperforms the other.
	Gursoy et al. explain that Ngram prunes sparse entries which leads to a spatial distribution that overestimates frequent intermediate regions and the remaining regions become excessively sparse.
	
	According to Deldar and Abadi, their model DP-MODR provides better results than AdaTrace for GeoLife and about similar results for Porto Taxi,
    though, they do not mention that their results for AdaTrace highly deviate from Gursoy et al.'s 
    evaluations on the same dataset (GeoLife)\footnote{It should be noted that the preprocessing is 
	not entirely equal, yielding 17,000 trajectories in Deldar and Abadi's work and 14,650 in Gursoy's paper, but it is not clear, whether this can explain the substantial difference in performance. In addition, Gursoy et al. do not specify their grid resolution which is another potential reason for the observed inconsistency.} and similarity 
	measures (location popularity ranking, trip error, and 
	diameter error). For example, for $\epsilon$ = 0.5, Gursoy et al. obtain a trip error of 0.048 for AdaTrace, while Deldar and Abadi report a value of 0.61 for AdaTrace and a value of 0.26 for DP-MODR. As the trip error is based on JSD, a smaller value means higher similarity, thus, according to one evaluation, AdaTrace would be superior, according to the other DP-MODR. We found similar inconsistencies in the location popularity ranking and the diameter error.
	Based on the provided evaluation it cannot be determined which approach outperforms the other.  
	The same is true for Liu et al. \cite{liu_adaptive_2019} who also state to provide a  model superior to AdaTrace but do not mention the inconsistency with Gursoy et al.'s results. Both use the same evaluation measures (query error, length error, and diameter error) and the same dataset  Brinkhoff. 
	Even though they explicitly extend the existing Ngram model to consider OD distributions, they, unfortunately, do not include such an evaluation.

    PrivTrace outperforms AdaTrace and DPT on three datasets in terms of diameter, length, and spatial distribution, as well as travel patterns according to a set of evaluation parameters, such as histogram bin size and grid size, which result in a higher granularity than those chosen in the AdaTrace paper.
    It should be noted that, unlike AdaTrace, PrivTrace does not construct origin-destination pairs of synthetic traces directly according to the raw dataset's distribution. Instead, the generation of a trace ends when a virtual end in the Markov chain is reached. Thus, an evaluation of the OD distribution in comparison to AdaTrace would have been desirable. 
    
    SPRT also outperforms DPT by far and AdaTrace slightly on various measures evaluating mobility characteristics on OD, diameter, length, spatial distribution, and travel patterns. Additionally, as their approach focuses on road network matching, they evaluate traffic violations, which they compute as the share of trajectories that deviate more than 20 meters from the road network. Unsurprisingly, the benchmarks fail with violations between $30\%$ to $50\%$, while SPRT only produces $4\%$ to $8\%$ of trajectories that violate traffic rules.

    \textbf{A set of models use deep learning architectures without any privacy guarantees:}
	Huang et al. \cite{inproceedings} combine a VAE and sequence-to-sequence (seq2seq) model to create \textit{SVAE} (Sequential Variational Autoencoder). The evaluation is limited to comparing the distance between synthetic trajectories and their real counterparts which is generally smaller than 800 meters. The authors also investigate the diversity of trajectories, pointing out that similarity and diversity oppose one another. It is noteworthy that the approach allows for different presentations of the locations, namely as coordinates, grid IDs, and a combination of the two. Their evaluations indicate that using coordinates only is the worst choice.
	
	Badu-Marfo et al. \cite{badu-marf_composite_2020} propose a model which is unique in including 
	demographics. The model is split into two components, a  tabular 
	demographics component, and a  sequence component, which are learned separately. In particular,  this implies that 
	demographics and trajectories are modeled separately. According to their evaluation, the distributions of trip lengths and route segment usages are well preserved, though no information is provided on how the 
	route segment usage is determined. According to the paper, it is to be assumed that staypoints of the used survey dataset are synthesized and a routing algorithm creates routes between the staypoints. Thus, the waypoints are not created by the proposed model, only OD pairs are. An analysis of OD distributions is not included.

	Wei et al. \cite{wei2020we} propose \textit{MoveSD} (human movement with system dynamics), which models agents' decision processes with a generator that learns a movement policy.
	The possible actions are limited to those that are defined ahead, i.e., turn right, left, u-turn, go straight or stop. The evaluation is based on the downstream task of next-location prediction. 
    The authors motivation is not directed at providing privacy, instead they aim to build a model of human mobility and synthetic trajectories are rather perceived a byproduct for evaluation purposes.
	
	\textit{TrajGAIL} (trajectory generative adversarial imitation learning) is proposed by Choi et al. 
	\cite{choi_trajgail_2021}. Although they motivate their work with privacy issues, they do not include any 
	privacy guarantees or analyses in their work. Like \cite{wei2020we}, possible actions are restricted to a 
	provided action set. Unlike previously stated models that are based on mapping points to grid cells, 
	TrajGAIL maps coordinates to a road network and can thus provide synthetic data which is in 
	accordance 
	with the road network instead of centroids of grid cells. On the downside, it is only designed for a 
	chessboard-like road network. More actions could be defined, though this would be non-trivial for diverse 
	road network layouts. As Cao et al. \cite{cao2021generating} point out, TrajGAIL bases the decision 
	for the next location on the current location only.
	
	\textit{TSG} (Two-Stage GAN) by Wang et al. \cite{wang_large_2021} uses one GAN to capture spatial 
	patterns of trajectories on a grid and a second to fine-position these to ensure realistic coordinates on the road network. Thus, unlike most grid-based approaches, TSG explicitly includes road mapping. 
	The evaluation includes screenshots of selected road segments that suggest a superiority in road network accuracy compared to other approaches. Unfortunately, they did not quantify road network accuracy with a similarity measure. 
	The evaluation also lacks OD and travel pattern analyses.
	
	Cao et al. \cite{cao2021generating} present \textit{TrajGen}, where spatial and temporal information are 
	learned in two separate steps: the spatial information learning is formulated as an image generation problem by 
	mapping trajectories to image pixels. Points on generated images are matched onto road segments of a 
	given map (i.e., OpenStreetMap) which are used to create a sequence with a 
	sequence-to-sequence model. Thus, like \cite{wang_large_2021}, they include map matching for 
	plausible 
	trajectories in terms of a road network.
	The start time of each trajectory is determined with an artificial neural network that takes the 
	length and the first location of a trajectory as input and outputs the time slot with the highest 
	probability. Thus, it is one of the few `trips' models that considers the temporal component.
	According to their evaluation, the spatial distribution per hour of day is well maintained. It is, however, noteworthy that 
	the evaluation is conducted on a taxi dataset where each trajectory spans the \textit{entire} daily ride of a taxi, with a 
	median of $307km$ per trajectory. Thus, as only the start time of the entire trip is 
	generated according to the spatial distribution, it is unlikely that this would correctly reproduce 
	spatio-temporal patterns over the entire day considering this data input. The authors do not visualize or quantify the deviation of the spatial distribution over the course of a day, therefore it is possible that the spatial patterns do not change a lot and those are learned by the mode. It would be interesting to extend the  
	evaluation to other measures and datasets. Also, single taxi rides might be better suited than daily taxi trajectories. 
     Xiong et al. \cite{xiong_trajsgan_2023} developed \textit{TrajSGAN} which is the first to model travel mode and purpose, using a semantic-guiding GAN with a CNN discriminator to determine how `real' the generated 2-D image appears. TrajSGAN outperforms LSTM-TrajGAN and DeltaGAN based on the MTL Trajet dataset of GPS trips, though we consider the two benchmark approaches part of the `user movements' category and thus not necessarily suited for this type of data.
    Jiang et al. \cite{jiang_continuous_2023} propose \textit{TS-TrajGen}, which aims to provide spatial continuity so that points do not `jump' in space as well as match the road network. Specifically, a two-stage GAN is built, based on prior domain knowledge of human mobility integrated with model-free learning paradigm. According to their evaluation, they outperform SVAE, MoveSim, TSG and TrajGen by far, covering mobility characteristics of distance, radius of gyration, location frequency and per-instance similarity measures. Though, the very small per-instance errors, which compare raw trajectories with synthetic counterparts with similar origin and destination, gives rise to privacy concerns, especially as a privacy evaluation is lacking.
	
	\textbf{Only two approaches provide differential privacy guarantees in combination with a deep 
	learning 
		stack}:
	Yu \cite{yu_reconciling_2019} proposes \textit{DeepSynthesizer}, an LSTM architecture that includes a start and stop symbol within the vocabulary to enable the model to learn distributions of origins and destinations. 
	\textit{DP-Loc} (Differentially Private Synthetic Trace Generator) was recently proposed by Lesty{\'a}n et al. \cite{lestyan_search_2022}, which also 
	includes the temporal component of trajectories:
	a VAE is used to generate OD pairs together with a start time. Transition 
	probabilities are learned with a feed-forward network to approximate the distribution of all paths 
	between the source and destination (at a given time). 
	They include extensive evaluations and compare their results with Ngram 
 and AdaTrace, finding that DP-Loc 
	outperforms AdaTrace on all similarity measures while Ngram is superior in terms of spatial density and similar with respect to frequent patterns. It should be noted that the trip length is only evaluated in terms of the number of points per sequence and not according to the actual distance. 
	
	\textbf{General considerations: } 
    The spatial granularity of trip data is a relevant utility consideration from a practitioner's perspective.
    For example, a resolution according to $500 m x 500m$ cells might be sufficient to identify highly frequented neighborhoods, but too coarse for tasks that require mapping to the street network, like determining which streets are preferred routes of cyclists.
    Using differentially private mechanisms, high dimensionality typically comes with high levels of noise which also applies here: according to the analysis of \cite{gursoy_utility-aware_2018-1}, the utility decreases due to increased noise, for finer grids than $11x11$ cells and \cite{du_ldptrace_2023} even find such an increase for grids finer than $6x6$ for three of their evaluated datasets\footnote{The exact numbers depends on the choice of the dataset, the spatial bounding box and the amount of privacy budget, though the general issue of finer grids remains in this context.}.
    As the covered area highly varies depending on the city the dataset origins from, which ranges from small towns like Oldenburg to huge cities like Beijing within the evaluations, it is difficult to translate the grid resolution into kilometers. Though, to obtain an idea of the scale, let's consider an area of $15 km$ x $15 km$ which roughly covers the denser part of a large city like Berlin but not even nearly the area of the GeoLife dataset in Beijing. It would need $30 x 30$ of $500 m$ cells, and grid sizes of $11x11$ ($6x6$, respectively) would signify a cell edge length of $1.4 km$ ($2.5 km$), questioning the usefulness of such trip data.
    This utility-privacy trade-off is tackled in different ways:
    Adaptive grids are used by AdaTrace and PrivTrace. DPT and TGM derive the grid resolution based on the dataset's sampling rate. LTPTrace, TGM, TrajGAIL, and SPRT limit the number of possible transitions to neighboring cells, thus, the dimensionality of the transition matrix is immensely reduced.
    DP-Loc only uses the top visited cells that cover 95\% of all visits and drops the remaining cells.
    
    SPRT additionally addresses the issue of plausibility with regard to matching the street network, by removing all implausible cells and transitions from the transition matrix. Without considerations of privacy guarantees, street mapping is further addressed by the following approaches: reinforcement learning implementations like \cite{wei2020we, choi_trajgail_2021} use street segments as location representations, but they are limited to a pre-defined action set (i.e., `left', `right', etc.), thus they would require additional adjustments to be able to map non-uniform road networks. 
    Two GAN \cite{wang_large_2021, xiong_trajsgan_2023} approaches use a CNN-based discriminator to learn plausible images, TrajGen \cite{cao2021generating} makes use of OpenStreetMap to map locations onto the closest road segments, and TS-TrajGen \cite{jiang_continuous_2023} uses an adjusted A* path search algorithm to ensure spatial continuity and includes topological constraints to enforce road matching.

    Table \ref{tab:mobCharac} provides an overview of mobility characteristics considered for statistical similarity evaluations. 
    Next to the spatial distribution, relevant mobility characteristics for trips include OD flows, trip length, and the chosen route, i.e., the travel pattern. Accordingly, many of the coded publications within this category consider respective similarity measures.
     As many models for trips omit the temporal dimension, we rarely see evaluations of the (spatio-)temporal mobility characteristic, although this would entail relevant information such as traffic volume per hour of day or average speed per road segment. 
    User patterns are not considered in any of the corresponding works; they are also likely not relevant for many use cases that consider trip data.

 \renewcommand*{\arraystretch}{1.2}
    \begin{footnotesize}
    \begin{longtable}[tb]{p{24mm}p{30mm}p{28mm}p{20mm}p{14mm}}
    	\caption{Overview of mobility characteristics that are evaluated (through similarity measures) by respective publications in each category.	\label{tab:mobCharac}} \\
    \hline
     \centering 
     \textbf{Mobility \newline characteristic} & \textbf{Trips} & \textbf{User movement} & \textbf{City population} & \textbf{Unspecified  category}  \\
     \hline
    spatial distribution & \cite{gursoy_utility-aware_2018-1, gursoy_differentially_2019-1,deldar_enhancing_2020, liu_adaptive_2019, wang_large_2021, cao2021generating, badu-marf_composite_2020, sun_synthesizing_2023, du_ldptrace_2023, wang_privtrace_2022} &
      \cite{bindschaedler_synthesizing_2016-1, zhao_synthesizing_2019, ouyang_non-parametric_2018-1, xusimulating, feng_learning_2020, mir2013dp,tamura_synthetic_2022} &
      \cite{pappalardo_data-driven_2018-1, bwambale_getting_2021} &
      \cite{chen2012differentially, chen_differentially_2012-3} \\
      \hline
      trip length &
      \cite{gursoy_utility-aware_2018-1, gursoy_differentially_2019-1, wang_large_2021, deldar_enhancing_2020, liu_adaptive_2019, kang_trag_2021, badu-marf_composite_2020, lestyan_search_2022, he_dpt_2015-1, ghane_tgm_2020, yu_reconciling_2019, sun_synthesizing_2023, du_ldptrace_2023, wang_privtrace_2022, jiang_continuous_2023} &
      \cite{feng_learning_2020, xusimulating} &
      \cite{berke_generating_2022, smolak_population_2020,pappalardo_data-driven_2018-1, anda_synthesising_2021} &
       \\
       \hline
       temporal distribution &
      \cite{kang_trag_2021, ghane_tgm_2020} &
      \cite{rao_lstm-trajgan_2020-1,tamura_synthetic_2022} &
      \cite{pappalardo_data-driven_2018-1, anda_synthesising_2021} &
       \\
       \hline
        spatio-temporal \hspace{0.5cm} distribution &
      \cite{lestyan_search_2022, cao2021generating} &
      \cite{ouyang_non-parametric_2018-1, xusimulating, rao_lstm-trajgan_2020-1, bindschaedler_synthesizing_2016-1, feng_learning_2020,tamura_synthetic_2022} &
      \cite{smolak_population_2020,anda_synthesising_2021, pang_development_2020, berke_generating_2022} &
       \\\hline
    OD flows &
      \cite{gursoy_utility-aware_2018-1, gursoy_differentially_2019-1, he_dpt_2015-1, deldar_enhancing_2020, yu_reconciling_2019, lestyan_search_2022, cao2021generating, sun_synthesizing_2023, du_ldptrace_2023, jiang_continuous_2023} &
       &
      \cite{berke_generating_2022} &
       \\\hline
    travel patterns &
      \cite{gursoy_utility-aware_2018-1, gursoy_differentially_2019-1, choi_trajgail_2021, he_dpt_2015-1, ghane_tgm_2020, yu_reconciling_2019, lestyan_search_2022, deldar_enhancing_2020, lestyan_search_2022, sun_synthesizing_2023, du_ldptrace_2023, wang_privtrace_2022} &
       &
      \cite{anda_synthesising_2021} &
      \cite{chen_differentially_2012-3} \\\hline
    
    user patterns &
       &
      \cite{feng_learning_2020, xusimulating, zhou_toward_2021-1} &
      \cite{pappalardo_data-driven_2018-1, berke_generating_2022, anda_synthesising_2021} & \\
    \end{longtable}
    \end{footnotesize}

	\subsection{User movements}
	\label{uc:user_movements}

A different perspective considers mobility data as a sequence of stay locations of a user, potentially over a longer period of time. Commonly considered datasets are LBSN data (e.g., Foursquare) or 
	mobile phone GPS traces. 
	Realistic datasets should entail users that re-visit the same locations multiple times, and travel distances 
	between locations should be reasonably small. The palette of algorithms for user movements is 
	diverse, especially in terms of deep learning approaches. Many are directed at a specific type of mobility 
	dataset or application. The heterogeneity of applications and datasets makes it rather difficult to 
	compare the approaches with one another.
	Specifically, useful similarity measures include the activity space (e.g., radius of gyration), number of daily 
	visited locations per user, I-Rank, or stay durations per location.

	Bindschaedler and Shokri \cite{bindschaedler_synthesizing_2016-1} published a model in 2016, 
	commonly referred to as \textit{SGLT}, which has been widely known and cited since (197 citations on 
	Google Scholar, 06.03.2023).
	For each user, represented by a single trajectory in their approach, a mobility model is computed that encompasses their spatio-temporal behavior (e.g., speed or duration of visit) and all models are combined into an aggregate mobility model. 
	A seed trace, i.e., a trajectory from the original data, is transformed into a `semantic seed' by replacing each location with their semantic class, such as `home'. The semantic seed in combination with the aggregate mobility model provides the base to create a synthetic trajectory that is semantically and geographically plausible.
	The authors provide the privacy guarantee of plausible deniability.
	As \cite{gursoy_utility-aware_2018-1} showed, SGLT is extremely slow which aligns with the authors' statement that 
	one trajectory takes about two minutes to be generated and thus does not scale well for any larger 
	dataset (only 30 user trajectories are used within their evaluation).

	Dandekar et al. \cite{dandekar_trajectory_2016} use smart card data, modeled as staypoints, to simulate trajectories in 
	communities of commuters. The communities are identified by clustering and for each community, a weighted directed graph is built to generate synthetic trajectories.
	
	Zhao et al. \cite{zhao_synthesizing_2019} choose a rather different approach: they propose 
	\textit{W$^3$-tess}
	which synthesizes privacy-preserving traces by enhancing the plausibility of synthetic traces by incorporating information from social networks. They argue that friends have a strong influence 
	on one’s mobility, for example, meeting at a restaurant and social locations where users are geographically in contact with their friends are included in their model.
	This is an interesting approach as it tries to capture the complexity of people’s behavior beyond home and work locations. They compare their approach to SGLT \cite{bindschaedler_synthesizing_2016-1} and find that W$^3$-tess better preserves social behavior. In terms of spatial distribution they perform highly similar, W$^3$-tess showing slightly better results. Even though the temporal dimension is part of the model, there is no information on how it is operationalized (likely similar to time windows in SGLT), and there is no evaluation of the temporal distribution.
	Rao et al. \cite{rao_lstm-trajgan_2020-1} propose \textit{LSTM-TrajGAN}, 
	an  LSTM-based GAN, which, unlike most other approaches, 
	does not discretize the geodata input but instead models location coordinates as continuous variables. 
    Additionally to coordinates, LSTM-TrajGAN takes the attributes user ID, weekday, hour of day, and location category as input. The location category refers to point-of-interests, thus staypoints, such as `food', `gym', `outdoors or recreation' in the utilized dataset.
	Unlike most other approaches, the weekday is also considered as temporal information next to the hour of day.
	Spatial similarity is only computed on a per-instance level; statistical similarity is only evaluated for temporal properties. It is worth noting that the generator in LSTM-TrajGAN uses not only a noise vector but also original trajectories as input, thus trying to mimic a specific trajectory when generating a synthetic one. In particular, the resulting synthetic trajectory has the same length as the original one. It is further worth noting that in the generation phase, the model requires the original data to produce synthetic ones. (It is thus worth discussing whether the model classifies as a synthetic mobility data generation model in the sense of our definition.) 
	
	Similar to Cao et al. \cite{cao2021generating}, Ouyang et al. \cite{ouyang_non-parametric_2018-1} propose a GAN based on a 
	CNN that represents trajectories as images. Each pixel contains information about the location, the start time and duration of the visit.  A pixel can be visited up to $k$ times by an individual (i.e., there are multiple layers of the same 2D space). Unlike other approaches, this model especially puts a focus on stay durations and includes evaluations of the spatio-temporal distribution. Note that this model consequently requires input data entailing stay durations. 
	The publication does not provide information on whether it ensures strictly increasing arrival times within the trajectory.
	Additionally, even though the authors 
	emphasize the importance of the semantic sequence and point out that their model captures the entire 
	trajectory for generation, there is no utility evaluation of travel patterns. 
	
	Xu et al. \cite{xusimulating} propose \textit{DeltaGAN}, another GAN-based approach for stay locations which especially focuses on continuous time, meaning, strictly increasing arrival times of events happening at irregular time intervals, 
	and time-conditioned location generation, thus, creating location visits depending on the time of day. Among other benchmarks, it compares itself to Ouyang \cite{ouyang_non-parametric_2018-1} 
	and outperforms by far in terms of total trajectory distance per day, radius of gyration, and number of unique locations per daily trajectory, all of which are different measures of user patterns. It also performs better on the same measures proposed by Ouyang et al., namely, the total stay duration of each location and the visit probability of each location (per time unit).

	Feng et al. \cite{feng_learning_2020} established the GAN framework \textit{MoveSim}. 
	The generator consists of two parts: a model-free 
	(i.e., not based on prior knowledge) self-attention mechanism and a model-based part, which provides information about the physical distance of locations, function similarity (which refers to POI category distribution), and historical transition matrix between all locations.
	To further capture important mobility regularities, the discriminator includes a correction term to regulate temporal periodicity and spatial continuity. Temporal periodicity describes that people tend to visit the 
	same locations at the same hour of the day, and spatial continuity encourages the model to limit travel distance.
	Like DeltaGAN, they benchmark with Ouyang et al. and show better results for user patterns, specifically total trajectory distance per day, radius of gyration, I-Rank, and the number of unique locations per daily trajectory. They also outperform Ouyang et al. on the stay duration per location and the overall location ranking.

	Zhou et al. \cite{zhou_toward_2021-1} created \textit{STULIG} 
	(Semi-supervised Trajectory-User Linking model with 
	Interpretable representation and Gaussian mixture prior) which is desgined for check-in data. The paper is rather focused on tackling the 
	trajectory-user linkage issue and thereby also generates synthetic trajectories.

	Tamura et al. \cite{tamura_synthetic_2022} propose \textit{Agent2Vec}, an approach inspired by Word2Vec. User trajectories are divided into 30-minute time intervals with corresponding locations. Locations are clustered using $k$-means with $k=4$ to represent the clusters 
	`residential', `office', `restaurant' and `other'.
	A Word2Vec approach creates for every user a vector representation that 
	captures the user's tendency to move and stay (over the course of day).  Synthetic data is created based on the clusters, stay 
	densities in cells, and travel distance distributions. According to the authors' evaluation, their approach still needs improvements in terms of maintaining travel distances.

    Chiesa and Taraglio \cite{chiesa_2022} focus on staypoints and stop durations using a VAE model. They neither benchmark against other models nor include similarity measures, instead, they rely on visual comparison of distributions of real and synthetic data.
	
	As mentioned at the beginning of this section, the proposed models for user movements are rather heterogeneous in their approaches and thus difficult to compare: some focus on a specific type of dataset, like CDR data \cite{mir2013dp}, a specific mobility topic, such as commuting patterns \cite{dandekar_trajectory_2016} or stay durations \cite{ouyang_non-parametric_2018-1}, or include specific aspects, e.g., social behavior \cite{zhao_synthesizing_2019}.
	Only three \cite{mir2013dp, bindschaedler_synthesizing_2016-1, zhao_synthesizing_2019} provide privacy guarantees, two other approaches include a privacy 
	evaluation \cite{rao_lstm-trajgan_2020-1, zhou_toward_2021-1} while the rest of the publications does 
	not further consider privacy. 
	Time is considered by almost all publications to some degree and many include evaluations concerning the spatio-temporal distribution (see Table \ref{tab:mobCharac}). Unfortunately, some only provide little information on how time is operationalized. 
    Only three publications consider user patterns and none evaluate travel patterns in terms of frequent sequences. This is striking as maintaining such patterns should be a central element for models in this category, which is also argued by many of the authors. Thus, for model validation, a respective evaluation would be helpful. If it is only of interest to maintain the spatio-temporal characteristic in the generated dataset, essentially the information, how many people are at a location at a certain time, the effort of creating plausible \textit{sequences} could be omitted.
	
	\subsection{City population} 
	\label{uc:day_population}
	Unlike synthesizing a single (likely not representative) mobility dataset, the goal of this setting is to create representative mobility data for a city population over the course of a day (as stay trajectories), for example as input for traffic models. 
	Mobility datasets are often just one piece of information used in 
	those approaches. Additional information include census data and models about mobility behavior like the exploration and preferential return mechanism \cite{song_modelling_2010}. Also, many models rely on the deduction of home and work 
     \cite{ berke_generating_2022, bwambale_getting_2021, mir2013dp,pappalardo_data-driven_2018-1} locations and make assumptions about commuting behavior. Thus, it should be noted that the performance of these models likely differs for unusual mobility behavior, for example, during the COVID-19 lockdowns.
    As elaborated in Section \ref{sec:privacy}, many of these models do not motivate their work with privacy concerns but rather stem from the research field of traffic modeling.
    Thus, only one includes differential privacy guarantees \cite{mir2013dp}, two coded publications include a privacy evaluation \cite{smolak_population_2020, berke_generating_2022} and one further publication uses privacy to motivate their work \cite{anda_synthesising_2021}.
	
	Time is commonly considered as fixed time windows and a location is assigned for each window and 
	user. While approaches for user movements could potentially work with temporally sparse 
	data, or such that do not include all types of locations (e.g., LBS check-in datasets might not 
	entail home or work locations), this would not suit the aim of the city population category. Original datasets usually consist of mobile phone or household survey data.
		
	Mir et al. \cite{mir2013dp} proposed \textit{DP-WHERE} in 2013, which has been widely cited since (157 
	citations on Google Scholar, 06.03.2023). It modifies the algorithm WHERE (Work and Home Extracted REgions) 
	\cite{isaacman_human_2012} by adding noise to achieve differential privacy. WHERE and DP-WHERE are  
	tailored to call detail record data and construct cumulative distribution functions as a base to generate new trajectories, using distributions of home locations, commuting distances per home region, work locations, distribution of calls in a day, probabilities of a call at each minute of a day, probabilities of a call at each location per hour.
	As mobile phone usage has increased over the past 10 years, CDR data is less sparse, producing temporally 
	more fine-granular trajectories. 
	Thus, this approach might not be suited anymore to generate CDR data as the single records are limited to work and home locations of users. When considering this model its suitability for more fine-granular data should therefore be tested previously. 

    \textit{3W} (WHO-WHERE-WHEN) follows the same idea of assigning synthetic persons a home and work location and sampling points according to temporal patterns. Though, instead of only considering home and workplace as locations, they define an action space around these two locations, and for every hour for a one-month-long period locations are sampled, either `home', `work' or a random sample from the action space.
	Privacy is evaluated by comparing the similarity between every trajectory pair.
  As there is no dependency between consecutive points, it is not surprising that the authors find the approach needs further development to also reproduce individual characteristics of human mobility in addition to characteristics on a population level. Bwambale et al. \cite{bwambale_getting_2021} create a synthetic population based on household and census data. Then, travel survey data and aggregated CDRs are used for trip generation modeling.

  The following models consider dependencies in the sequence of locations. Papparlado and Simi \cite{pappalardo_data-driven_2018-1} propose \textit{DITRAS} which works in two steps: first a diary is generated (temporal pattern) with abstract locations based on a 
	Markov chain (non-parametric), secondly, the abstract locations are replaced with geographical 
	information using a weighted spatial tessellation and an exploration and preferential return model 
	(parametric). The authors state that exploration and preferential return comes with the downside of overestimating long-distance trips, also they envision the usage of more complex typical diaries. 
	\textit{Digital Twin Travellers} \cite{anda_synthesising_2021} is also based 
	on Markov models and aims to recreate realistic daily schedules of sequences of staypoints.
	They benchmark against DITRAS \cite{pappalardo_data-driven_2018-1} which, according to their evaluation, is outperformed on all measures: start time distribution, duration distribution, number of trips, 
	frequent patterns (`tour network'),  spatial error, distance traveled, activity space, and mobility entropy.
	Pang et al. \cite{pang_development_2020} (also \cite{pang_modeling_2017, pang_replicating_2018}) 
	use a reinforcement learning approach and additionally include information on context features, namely the number of offices, the number of employers, the number of schools, the number of evacuation facilities, the number of amusement facilities, the length of roads, railway stations, and the residential 
	density. They use fixed time intervals which can span only seconds or longer periods -- within their evaluation 30 minute intervals are used, focusing rather on staypoints than waypoints.

	Berke et al. \cite{berke_generating_2022} use a text-generation-based model in their recent proposition, putting locations into analogy with letters in a text. They discretize time and geographic space with one-hour time windows and census cells, respectively. Home and work locations are deduced from the data and used as input to train an RNN that generates a cell for each time window. 
 Null values allow the representation of missing values. Either the raw data distribution of home and work locations or publicly available census data can serve as input to 
	generate synthetic data. The utility evaluation is rather short, only comparing trip distance, locations per user, and the proportion of aggregate time spent per location. Privacy is evaluated by measuring the distance between any two trajectories.

	For this category, it is especially relevant to maintain the spatio-temporal distribution, thus this is evaluated by 4 of the 7 publications. Additionally, trip length and user patterns are evaluated by a subset of publications.
    Generally, the authors of \cite{pappalardo_data-driven_2018-1, smolak_population_2020, anda_synthesising_2021} include extensive evaluations and provide detailed information on (dis)advantages. Bwambale et al. \cite{bwambale_getting_2021} and Pang et al. \cite{pang_development_2020} require specific additional user input, which makes them only suitable if the specific prerequisites are given. For an in-depth assessment of Berke et al. \cite{berke_generating_2022}, further evaluations are needed. A comparison between their RNN approach and the other approaches of this category would provide interesting insights.
	
	\subsection{Unspecified category}
	\label{uc:arbit_sequence}

	Next to Chen et al. \cite{chen2012differentially,chen_differentially_2012-3} (as already introduced earlier), all other publications within this category are generic or do not provide enough information to classify them according to one of our categories. 
  For example, \cite{kulkarni_generating_2017-1, kulkarni_generative_2018-1, yu_reconciling_2019} use generic RNN architectures that can potentially take any arbitrary sequence as input.
	
	Li et al. \cite{li2016differentially} propose a model which first applies a trajectory generalization step by 
	grouping all spatial points at each timestamp with $k$-means clustering. To provide differential privacy, it uses the exponential mechanism to select the partitions when constructing new trajectories, and finally adds Laplace noise to the counts of trajectory sequences. $k$-means requires to pre-define the number of clusters, though the authors do not elaborate on how this can be set 
	mindfully. For their evaluation, $k$ is set to 50 which appears to be a rather coarse setting. The short evaluation limits the possible conclusion on the model's utility. 
	Chen et al. \cite{chen2020rnn} propose \textit{RNN-DP}, an RNN-based approach which is mainly directed at an entirely 
	different scenario of handling real-time trajectory data instead of the full historical data. Still, they 
	include a section about trajectory release which basically follows the same clustering approach as 
	\cite{li2016differentially}.
	
	Sakuma et al.'s \textit{CANDAR} \cite{sakuma2021trajectory} uses a Seq2Seq autoencoder model that adds differentially 
	private noise in the latent space. Next to \cite{rao_lstm-trajgan_2020-1, inproceedings}, this model is the 
	 one of the few that does not discretize locations but instead keeps a continuous representation. The GPS smartphone data used for evaluation is preprocessed by grouping it per user and day, then
	trajectories with less than 5 points are removed, and those with more than 10 points are split again, thus the maximum 
	length of a trajectory is 10. Given this preprocessing, it is difficult to determine a suitable real-life dataset that would match these data characteristics.
	
	Kulkarni and Garbinato \cite{kulkarni_generating_2017-1} propose a basic LSTM model that takes 
	a sequence of discretized locations in addition with selected time-series features (which are not further 
	specified) and outputs a sequence of locations. According to the authors, the ordering of commonly visited places is \textit{not} well preserved by their model.
	Generally, RNNs could be used for any type of trajectory, though this implementation is not well suited for the category \textit{trips}, as 
	they use a pre-defined fixed length sequence as input and output, thus neither distributions of OD pairs nor trip lengths are likely to be well maintained. Also, there is no information on how the initial location to initiate the 
	generation process is determined. Using the actual initial locations is a privacy issue while taking random locations is a utility problem. Blanco-Justicia et al. \cite{blanco-justicia_generation_2022} propose a similar model based on a bidirectional LSTM, thus the same shortcomings apply.
	To ensure that the model does not reproduce original trajectories too closely, they collect the top-$k$ predictions for the next point and choose one of them uniformly at random. In 2018, Kulkarni et al. \cite{kulkarni_generative_2018-1} also compared different implementations of 
	RNNs, GANs, and copulas, though providing only limited information on model descriptions and evaluations. 
	
	Zhan \cite{zhan2022privacy} recently proposed \textit{LSTM-PAE}, an LSTM-based location privacy protection mechanism via representation learning and adversarial learning, to learn a privacy-preserving feature extraction encoder. They focus on the trade-off between utility (measured as the accuracy of 
	next-location prediction) and privacy (measured as re-identification risk). While such a detailed comparison is an appreciated approach, the paper lacks information on data preprocessing such as how discrete locations are obtained for the one-hot encoding or how trajectories are transformed into a pre-defined sequence length. Even though the model takes one-hot-encoded timestamps as input, the time representation of the synthetic data remains unclear.
	The utility is only evaluated in terms of prediction accuracy.
	
	Alatrista-Salas et al. \cite{10.1007/978-3-031-13448-7_7} propose a differentially private GAN and compare this to a regular GAN, only providing little information about the model and its evaluation. PrivTC \cite{yang_2022} propose a local differentially private model without requiring specific trajectory semantics. It is evaluated with the Gowalla LBSN dataset as well as a taxi dataset and outperforms Ngram based on query error, frequent pattern similarity and distance error.
    
 \section{Discussion and Conclusion}
\label{sec:discussion}

Generating synthetic urban mobility data holds much potential, but it is a complex task with no one-size-fits-all solution that could serve this heterogeneous field. This is also reflected in the diversity of the reviewed models. However, the absence of a clear definition for high utility in synthetic datasets makes model comparison challenging.
There are no straightforward downstream tasks suited for evaluation; instead, mobility data is used for a variety of non-standardized analyses and models that pose different requirements to the data, reflected by different statistical distributions, potentially entailing complex interactions.
Thus, a single valid definition of high utility can hardly be established. In addition, varying dataset formats make standardized evaluations even more difficult. For example, a temporal representation is either omitted completely, modeled as a start timestamp, or based on fixed time windows. 

It is likely impossible to create a model that preserves all distributions equally well while also maintaining privacy.
We suggest that researchers developing new models clearly state their intended use cases and provide meaningful evaluations covering all relevant mobility characteristics with at least one measure each.
Details on implementations of utility evaluations would be desirable, such as utilized grid resolutions for spatial aggregations and histogram bin sizes for all other aggregations. 
Such choices are often not stated or can only implicitly be deduced from the implementation.
This facilitates the results' interpretation, as the utility can easily be increased by the choice of a coarser resolution (for example, the spatial distribution might easily be maintained well on a $2 km$ x $2 km$ grid but not nearly as well on a $250 m$ x $250 m$ grid). Without respective details, practitioners do not have sufficient information to decide whether the reported utility satisfies their needs. A granularity standard can hardly be given in general as it highly depends on the intended use case.
Also, the provision of the model's source code and generated synthetic datasets would facilitate the evaluation and comparison with existing models.
Additionally, this field of research would benefit from real-world test cases of organizations that provide data and use synthesized trajectories for their actual purposes, which would reveal shortcomings in practical settings. 
This could also offer valuable insights into the necessary levels of similarity across different mobility characteristics in varying contexts.

While there are specific limitations for each model, there are (so far) also global limitations: many models entirely omit the temporal dimension. If included, it is only considered as the time of day (potentially also the day of the week). This means that the data does not reflect information on mobility behavior differences over the course of the year, e.g., seasonal differences, event-related mobility behavior (e.g., a sports event), or weather-related behavior adaptions.

None of the publications provides evaluations in relation to the dataset size. This is especially relevant for deep learning-based models that commonly need a lot of data to properly learn latent distributions.
Generally, not all publications provide sufficient information about the origin, format, size, and availability of the utilized datasets. 
Potential biases or dataset-specific characteristics are rarely stated by authors, even though they might impact the generalizability of the results. More careful considerations of trajectory semantics, such as the type (i.e., waypoints or staypoints) or the length of trajectories, and how many trajectories a user contributes, would be desirable, as this information is crucial for determining the suitability of models for specific contexts.

Privacy is one of the main motivations for synthetic data generation. Yet, many models lack privacy guarantees or privacy evaluations. 
As recent approaches tend to rely more on deep learning models, they also tend to omit privacy guarantees (see Figure \ref{fig:timeline}). This might not be surprising, as differentially private deep learning has only recently been provided as part of open-source libraries (e.g., Tensorflow Privacy in 2019\footnote{\url{https://github.com/tensorflow/privacy}}). Two deep learning models already make use of differentially private stochastic gradient descent methods \cite{yu_reconciling_2019, lestyan_search_2022} and likely more models will follow. 

Differential privacy is no silver bullet and it comes with certain limitations. Protecting the privacy of individuals leads to a utility reduction of data from minority groups \cite{pujol_equity_2021}. 
For example, a dataset might only include a small share of children who have a different mobility behavior than, say, commuters, who typically constitute a large subpopulation in mobility data.
In such a case, a differentially private synthetic dataset might well resemble commuters' mobility behavior but not the children's movements. 
It should be noted that deep learning models without any additional privacy guarantee mechanisms likely do not properly learn minority group behavior either \cite{blanzeisky_algorithmic_2021}. 
Additionally, differential privacy provides guarantees on the level of items (trips in this context) or users, and not on the level of groups. 
For example, the well-known incident of the released aggregated Strava data which revealed secret military bases \cite{noauthor_fitness_nodate} would not have been prevented by differential privacy. Thus, the publication of differentially private synthetic data still requires careful thought about the information that is still entailed.

In summary, for the contribution of meaningful new models, we consider the following aspects as valuable parts of the respective publication: Information about applicable use cases; a detailed description of the input and output data format of the proposed model;  datasets and configurations of measures used for evaluation; openly available source code as well as the created synthetic dataset(s). 
We see the potential for future research in the following aspects: A comprehensive comparison of different models 
based on a heterogeneous set of datasets and similarity measures would foster use case specific recommendations. 
Standardized benchmarking tools, such as the Synthetic Data Vault \cite{7796926} that provides a set of standard datasets and evaluations for tabular data or time series synthetic data generation models, could contribute to better comparability, accelerate the development of new models and create confidence in practitioners' decision processes. 
We have developed a Python package \textit{dp\_mobility\_report}\footnote{Package code repository: \url{https://anonymous.4open.science/r/dp_mobility_report-A35C/}} capable of computing a comprehensive set of measures for various mobility characteristics (so far only considering staypoints), allowing custom configurations of histogram bin sizes and tessellations. 
This could be extended to provide standard test cases comprising datasets and configurations.
Moreover, comparing deep learning and traditional methods would provide meaningful insights, with special regard to the achieved utility concerning the input dataset size.
 Finally, recent developments in the fusion of differential privacy and deep learning open the way for further research.

\bibliographystyle{ACM-Reference-Format}
\bibliography{references}

\section{Appendix}

\section{Links to Publically Available Model Source Codes}
\label{sup:datasets}

\begin{table}[h]
	\begin{tabular}{p{7cm}|p{6cm}}
		\hline\textbf{Model Name/Authors} & \textbf{Source Code Link} \\
		\hline
		AdaTrace \cite{gursoy_utility-aware_2018-1} & \url{https://bit.ly/3SEbdoG} \\
		DP-Star \cite{gursoy_differentially_2019-1} &  \url{https://bit.ly/3fmMYwR}\\
		TSG \cite{wang_large_2021} & \url{https://bit.ly/3SphvJ0} \\
		LSTM-TrajGAN \cite{rao_lstm-trajgan_2020-1} & \url{https://bit.ly/3Ckmpld} \\
		MoveSim  \cite{feng_learning_2020} & \url{https://bit.ly/3Sr5YZY} \\
		STULIG \cite{zhou_toward_2021-1} & \url{https://bit.ly/3SD6wvX}\\
		DITRAS \cite{pappalardo_data-driven_2018-1} & \url{https://bit.ly/3LRDaak}\\
		3W \cite{smolak_population_2020} & \url{https://bit.ly/3Sp4eAm}\\
		Berke et al. \cite{berke_generating_2022} & \url{https://bit.ly/3RrNDKF}\\
		Kulkarni and Garbinato \cite{kulkarni_generative_2018-1} & \url{https://bit.ly/3RBIVdL} \\
		DP-Loc \cite{lestyan_search_2022} & \url{http://bit.ly/3Zm3oHq} \\
		TrajGAIL \cite{choi_trajgail_2021} & \url{http://bit.ly/3EIgJ52} \\
		PrivTrace \cite{wang_privtrace_2022} & \url{http://bit.ly/3Zn65bW} \\
		LDPTrace \cite{du_ldptrace_2023} & \url{http://bit.ly/3mnYJ9P} \\
	\end{tabular}
\end{table}

\pagebreak

\section{Links to Publically Available Datasets}
\label{sup:sourceCodes}

\begin{table}[h]
	\begin{tabular}{p{7cm}|p{6cm}}
		\hline
		\textbf{Dataset Name} & \textbf{Download Link} \\
		\hline
		Bike NYC & \url{https://bit.ly/3E3P3rV} \\
		Bike Washington & \url{https://bit.ly/3zvH3wo} \\
		Boston Hubway Bike & \url{https://bit.ly/3y1rDjf} \\
		GeoLife (Beijing) &  \url{https://bit.ly/3Rqbjz8}\\
		Porto Taxi & \url{https://bit.ly/3RkGWtR} \\
		San Francisco Taxi & \url{https://bit.ly/40DaY1B} \\
		Microsoft T-Drive (Beijing 2011) & \url{https://bit.ly/3CjbYOE} \\
		Lausanne Data Collection Campaign (LDCC) & \url{https://bit.ly/3BXztvr} \\
		Foursquare Weekly & \url{https://bit.ly/3ULvhHv} \\
		NYC Restaurant Rich Dataset &  \url{https://bit.ly/3ULvhHv} \\
		NYC and Tokyo Dataset & \url{https://bit.ly/3ULvhHv} \\
		Brightkite & \url{https://stanford.io/3Cjc7BG}\\
		Gowalla & \url{https://stanford.io/3SDcT1K}\\
		Brinkhoff (Oldenburg) & \url{https://bit.ly/3y0YHYD}\\
		Survey Montreal & \url{https://bit.ly/3dUnWVx} \\
		MTL Trajet & \url{https://bit.ly/3IP35yt}
	\end{tabular}
\end{table}

\end{document}